\documentclass[acmsmall]{acmart}
\usepackage{bm}
\usepackage{multirow}
\usepackage[normalem]{ulem}
\usepackage{algorithm}
\usepackage{algpseudocode}
\usepackage{amsmath}
\usepackage{subfig}
\usepackage{float}
\AtBeginDocument{%
  }

\setcopyright{acmcopyright}
\copyrightyear{2018}
\acmYear{2018}
\acmDOI{XXXXXXX.XXXXXXX}

\acmJournal{TOIS}
\acmVolume{1}
\acmNumber{1}
\acmArticle{1}
\acmMonth{8}

\begin{document}

\title{Manipulating Visually-aware Federated Recommender Systems and Its Countermeasures}

\author{Wei Yuan}
\affiliation{%
  \institution{The University of Queensland}
  \city{Brisbane}
  \state{QLD}
  \country{Australia}
}
\email{w.yuan@uq.edu.au}

\author{Shilong Yuan}
\affiliation{%
  \institution{Nanjing University}
  \city{Nanjing}
  \state{Jiangsu}
  \country{China}
}
\email{shilongyuan@nju.edu.cn}

\author{Chaoqun Yang}
\affiliation{%
\institution{Griffith University}
  \city{Gold Coast}
  \state{QLD}
  \country{Australia}
}
\email{chaoqun.yang@griffith.edu.au}

\author{Quoc Viet Hung Nguyen}
\affiliation{%
  \institution{Griffith University}
  \city{Gold Coast}
  \state{QLD}
  \country{Australia}
}
\email{henry.nguyen@griffith.edu.au}

\author{Hongzhi Yin}\authornote{Corresponding author.}
\affiliation{%
  \institution{The University of Queensland}
  \city{Brisbane}
  \state{QLD}
  \country{Australia}
}
\email{db.hongzhi@gmail.com}

\renewcommand{\shortauthors}{Yuan et al.}

\begin{abstract}
  Federated recommender systems (FedRecs) have been widely explored recently due to their ability to protect user data privacy.
  In FedRecs, a central server collaboratively learns recommendation models by sharing model public parameters with clients, thereby offering a privacy-preserving solution.
  Unfortunately, the exposure of model parameters leaves a backdoor for adversaries to manipulate FedRecs. 
  Existing works about FedRec security already reveal that items can easily be promoted by malicious users via model poisoning attacks, but all of them mainly focus on FedRecs with only collaborative information (i.e., user-item interactions).
  We argue that these attacks are effective because of the data sparsity of collaborative signals.
  In practice, auxiliary information, such as products' visual descriptions, is used to alleviate collaborative filtering data's sparsity.
  Therefore, when incorporating visual information in FedRecs, all existing model poisoning attacks' effectiveness becomes questionable.
  In this paper, we conduct extensive experiments to verify that incorporating visual information can beat existing state-of-the-art attacks in reasonable settings.
  
  However, since visual information is usually provided by external sources, simply including it will create new security problems.
  Specifically, we propose a new kind of poisoning attack for visually-aware FedRecs, namely image poisoning attacks, where adversaries can gradually modify the uploaded image to manipulate item ranks during FedRecs' training process.
  Furthermore, we reveal that the potential collaboration between image poisoning attacks and model poisoning attacks will make visually-aware FedRecs more vulnerable to being manipulated.
  To safely use visual information, we employ a diffusion model in visually-aware FedRecs to purify each uploaded image and detect the adversarial images.
  Extensive experiments conducted with two FedRecs on two datasets demonstrate the effectiveness and generalization of our proposed attacks and defenses.

\end{abstract}

\ccsdesc[500]{Information systems~Recommender systems}

\keywords{federated learning, poisoning attack, multimodal recommendation, image pollution and purification}

\received{20 February 2007}
\received[revised]{12 March 2009}
\received[accepted]{5 June 2009}

\maketitle

\section{Introduction}\label{sec_intro}
Recommender systems have become an integral part of web applications (e.g., e-commerce~\cite{chen2020try,wei2007survey} and social media~\cite{yin2016spatio}), during the era of information explosion, since they are effective in reducing information overload by discovering users' potential interests.
Traditionally, recommender systems are trained in a centralized server using a vast collection of user data~\cite{zhang2019deep}.
However, with the growing awareness of privacy and the release of privacy protection regulations, such as the General Data Protection Regulation (GDPR)~\cite{voigt2017eu} in the European Union and the California Consumer Privacy Act (CCPA)~\cite{harding2019understanding} in the United States, collecting and storing user data has become more challenging.

Federated learning, as a privacy-preserving paradigm, allows for training models on decentralized data~\cite{mcmahan2017communication}.
Consequently, an increasing number of researchers are exploring the potential of federated learning in recommender systems, resulting in the emergence of federated recommender systems (FedRecs).
In FedRecs, the central server and users/clients\footnote{In this paper, client and user are equivalent, since a client is only responsible for one user considering privacy protection requirements.} collaborate to learn a recommendation model by sharing model public parameters instead of user private data.
Due to the significant advantage of data privacy protection, after the first FedRec framework proposed by Ammad et al.~\cite{ammad2019federated}, several extended versions have sprung up to enhance the effectiveness and efficiency of FedRecs~\cite{muhammad2020fedfast,liang2021fedrec++,imran2023refrs}.

Due to the exposure of model parameters to all participants, and some of them may have malicious intentions, the security issues of FedRecs have raised concerns among researchers.
Adversarial item rank manipulation is one of the most studied security problems in FedRecs, driven by financial incentives, which can lead to strong unfairness and even reduce the validity and usability of recommender systems.
In~\cite{zhang2022pipattack}, the first model poisoning attack, PipAttack, was introduced to demonstrate the vulnerability of FedRecs to being controlled by malicious participants who upload poisoned model updates.
After that, all existing works about item rank manipulation in FedRecs are based on model poisoning attacks.
For example, FedRecAttack~\cite{rong2022fedrecattack} argues that PipAttack requires too many malicious users, which is not practical.
It, on the other hand, achieves item promotion with fewer malicious users but requires more prior knowledge, such as a small proportion of user interaction data, which even violates the FedRec learning protocol.
~\cite{rong2022poisoning} proposed the first model poisoning attacks without any prior knowledge assumptions.
Nevertheless, its performance is unstable and undesirable because it randomly samples vectors from a Gaussian distribution to act as the proxy of user embeddings.
In our previous work~\cite{yuan2023manipulating}, we proposed PSMU, which achieves state-of-the-art attack performance without relying on any prior knowledge and with fewer malicious users and training epochs, revealing the severe threats of model poisoning attacks to FedRecs.

\begin{figure}[!htbp]
  \centering
  \includegraphics[width=\linewidth]{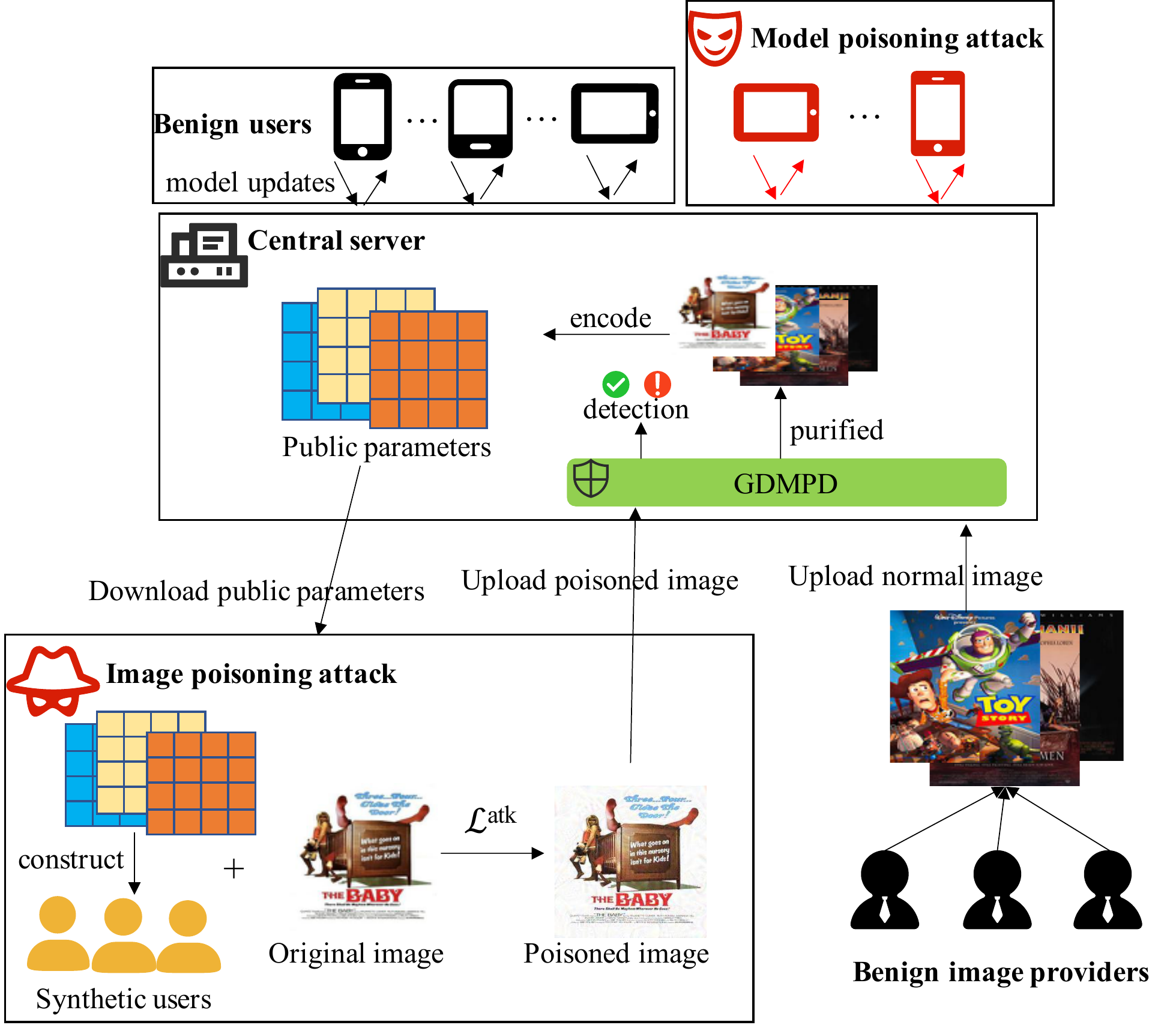}
  \caption{Overview of the new threats ``image poisoning attack'' and our diffusion model based defense mechanism.}\label{fig_image_poisoning}
  \Description{Overview of image poisoning attack.}
\end{figure}
However, all existing model poisoning attacks, including our previous work~\cite{yuan2023manipulating}, only verify the threat in FedRecs with collaborative data (i.e., user-item interaction data).
We argue that due to the inherent sparsity of collaborative information, many items, especially cold ones, lack sufficient descriptions.
Consequently, these attacks can easily manipulate the rank order of items by uploading poisoned gradients.
In real-life scenarios, item visual descriptions are used to alleviate collaborative data's sparsity problem.
Intuitively, incorporating these visual signals makes the item features more comprehensive and robust, and as a result, existing state-of-the-art model poisoning attacks may fail to promote items adversarially.
In this paper, we empirically show that all existing state-of-the-art model poisoning attacks fail the adversarial item promotion in visually-aware FedRecs. i.e.,
visual information can mitigate the adversarial promotion threat caused by model poisoning attacks.

While visual signals can alleviate the model poisoning problem, incorporating them may leave another backdoor for adversaries to manipulate item ranks, as the product visual descriptions are typically provided by external sources that are not always trustworthy.
In other words, the adversaries can be item image providers who promote the target items by uploading images with human-imperceptible noise.
In this paper, we refer to such attacks as \emph{image poisoning attacks}.
Fig.~\ref{fig_image_poisoning} presents an overview of image poisoning attacks.
It is worth noting that some research has used polluted images to change item ranks in centralized recommender systems (i.e., visual attacks~\cite{liu2021adversarial,cohen2021black}).
However, our image poisoning attacks have many different settings.
Specifically, in centralized recommender systems, the model parameters are not accessible unless a ``white-box'' assumption is made.
In contrast, the public parameters of FedRecs are apparent, but the user's private parameters are strictly out of reach.
Moreover, all research in centralized recommender systems~\cite{liu2021adversarial,cohen2021black} assumes adversaries can obtain benign users' feedback, which is not valid in FedRecs.
Furthermore, previous visual attacks for centralized recommendations can only be launched after the recommender system is well-trained, as adversaries cannot participate in the training process.
While the image poisoning attacks in FedRecs are continually executed during the training process.

In this paper, we propose the first image poisoning attack, namely PSMU(V) (\emph{p}oisoning with \emph{s}ynthetic \emph{m}alicious \emph{u}sers via \emph{v}isual information), to disclose the risk of directly using visual information in FedRecs as shown in the bottom left of Fig.~\ref{fig_image_poisoning}.
Specifically, PSMU(V) is an image poisoning version of our previous work PSMU which is a model poisoning attack.
The same as PSMU, PSMU(V) constructs a group of synthetic users with randomly selected interactions.
It then calculates image perturbations with attack objectives guided by these synthetic users. 
Finally, the adversary uploads the poisoned image to the central server to influence the target item's feature representation.
The above steps are iteratively executed with the training process of FedRecs.
Through experimental results, we demonstrate the effectiveness of our image poisoning attacks and reveal the risks of using images provided by external sources directly. 
Additionally, since PSMU and PSMU(V) can have a consistent attack objective and can be launched simultaneously, we propose PSMU++ (i.e., PSMU+PSMU(V)) to reveal a more severe threat caused by the potential collaboration of PSMU and PSMU(V).
That is, by launching both model poisoning attacks and image poisoning attacks, the target items will be more easily exposed to users than executing only one of these attacks.

The threats posed by image poisoning attacks underscore the need for a safer mechanism to use visual information.
However, the defense against image poisoning attacks is still under-explored.
In centralized recommendation, ~\cite{tang2019adversarial} attempted to employ adversarial training to improve visually-aware recommender systems' robustness, but it can only defend against untargeted attacks that aim to destroy a recommender system, and ~\cite{liu2021adversarial} indicated that adversarial training cannot effectively prevent item promotion attacks. 
Inspired by the great achievement of the Denoising Diffusion Probabilistic Model (DDPM)~\cite{sohl2015deep,ho2020denoising} in image generation, we propose our novel image poisoning defender, \emph{G}uided \emph{D}iffusion \emph{M}odel for \emph{P}urification and \emph{D}etection (GDMPD), as shown in the middle part of Fig.~\ref{fig_image_poisoning}.
GDMPD can achieve two functions: purification and detection.
The purification function aims to prevent adversarial images from achieving their malicious purpose.
Particularly, the purification is based on DDPM which includes two processes: diffusion process and reverse process.
During the diffusion process, the model gradually adds noise to the image, which can submerge the adversarial perturbations.
Then, the reverse process purifies these noises to recover the image, which can remove both added noise and adversarial perturbations.
In FedRecs, besides reducing the effectiveness of attacks, detecting malicious behavior is also necessary since it can provide the system manager with valuable insights for conducting further processes.
Therefore, our GDMPD provides the detection function to further indicate which image is adversarial.

To support the proposed attack and defense methods, we extend the base FedRecs (Fed-NCF and Fed-LightGCN) used in our previous work~\cite{yuan2023manipulating} to visually-aware FedRecs.
Then, we conduct extensive experiments with these FedRecs on two recommendation datasets (MovieLens-1M and Amazon Cell Phone).
The experimental results demonstrate that incorporating visual signals can alleviate model poisoning attacks but simply using visual information provided by untrusty sources will leave a backdoor for image poisoning attacks, and our novel defense method can fix such a backdoor.

To sum up, our major new contributions are listed as follows:
\begin{itemize}
  \item Our previous work~\cite{yuan2023manipulating} only studied the threat of model poisoning attacks for federated recommender systems on collaborative data. In this paper, we make the exploration of model poisoning attacks in visually-aware federated recommender systems. The empirical results demonstrate that visually-aware federated recommender systems are robust to existing state-of-the-art model poisoning attacks, since visual signals can alleviate the data sparsity problem of collaborative information.
  \item Although visual information can defend against model poisoning attacks, we propose the first \emph{image poisoning attack}, PSMU(V), to reveal a new backdoor for adversaries to promote items if visual information is directly used. To the best of our knowledge, this is the first work to reveal such threats in visually-aware FedRecs. Furthermore, we propose PSMU++ to investigate the potential hazard of collaboration between model poisoning attacks and image poisoning attacks.
  \item To fix the security hole of image poisoning attacks, we propose the Guided Diffusion Model for Purification and Detection (GDMPD), which is a diffusion model based defense mechanism in the central server of FedRecs to purify each uploaded image and detect the adversarial images. So far as we know, this is the first work that utilizes the diffusion model as defense method in federated recommender systems.
  \item We have performed comprehensive experiments using two visually-aware FedRecs that we extended from our previous work~\cite{yuan2023manipulating} on two widely-used recommendation datasets. The experimental results demonstrate the effectiveness and generalizability of our proposed methods. 
\end{itemize}

The remainder of the paper is organized as follows.
Related work is reviewed in Section~\ref{sec_related}, followed by the introduction of the visually-aware federated recommender systems in Section~\ref{sec_vfedrec}.
Section~\ref{sec_attack} presents the technical details of our attacks extended from our previous work~\cite{yuan2023manipulating}.
Then, in Section~\ref{sec_defense}, we show how to fix the security problem revealed by image poisoning attacks.
Section~\ref{sec_exp} exhibits a comprehensive analysis of experimental results.
Finally, Section~\ref{sec_conclusion} gives a brief conclusion of this paper.

\section{Related Work}\label{sec_related}
In this section, we briefly review the literature on four related topics: federated recommender systems, attacks and defense mechanisms for federated recommender systems, visually-aware recommender systems, and diffusion models.
Other involved topics such as the development of general recommender systems and federated learning can be referred to corresponding surveys~\cite{zheng2023automl,yu2022self,li2020federated}.


\subsection{Federated Recommender Systems}~\label{sec_fecrecs}
Federated Recommender Systems (FedRecs) have gained increasing attention in recent years due to their ability to protect user privacy.
Ammad et al~\cite{ammad2019federated} presented the first FedRec framework that applies federated learning with a collaborative filtering model.
Based on this basic framework, many extended versions have been proposed in a short time~\cite{alamgir2022federated,sun2022survey}.
Some works attempt to reduce the performance gap between FedRecs and centralized recommender systems.
For example, ~\cite{wu2021fedgnn} utilizes Graph Neural Network (GNN)~\cite{scarselli2008graph} to achieve fairly good recommendation accuracy.
Wu et al.~\cite{wu2022fedcl} employed contrastive learning in FedRecs.
Other works focus on the efficiency of FedRecs.
~\cite{chen2018federated,muhammad2020fedfast} investigate fast convergence of FedRecs, while Zhang et al.~\cite{zhang2022lightfr} proposed a lightweight communication strategy based on learning to hash (L2H)~\cite{wang2017survey}.
ReFRS~\cite{imran2023refrs} learns dynamic and diversified user preferences on resource-constrained devices.
Some work transplant FedRecs to specific recommendations, such as news recommendation~\cite{qi2020privacy}, social recommendation~\cite{liu2022federated}, POI recommendation~\cite{guo2021prefer}, and so on~\cite{lin2020fedrec}.

In addition to enhancing the effectiveness and efficiency of FedRecs, privacy concerns are also a crucial research direction in this area.
Chai et al.~\cite{chai2020secure} demonstrated that even if user data is not directly shared, adversaries can still recover sensitive information from the model updates sent by the target user.
To address this issue, a central server can apply a differential privacy (DP) mechanism to perturb the global model, as proposed in~\cite{wang2021fast}.
However, this approach assumes that the central server is in a sterile environment, which is not applicable in real-life scenarios where the server may be curious about clients' private information. 
To protect user privacy further, local differential privacy (LDP) is equipped on the client side~\cite{wu2021fedgnn}. 
Zhang et al.~\cite{zhang2022comprehensive} introduced adaptive LDP, which can protect privacy with less impact on recommendation performance.
Nevertheless, Yuan et al.~\cite{yuan2023interaction} discovered that LDP alone cannot safeguard user-item interaction information, which is known as the user-item interaction leakage problem.
They provided a regularization-based method to tackle such a problem.
Additionally, ~\cite{yuan2023federated} enables FedRecs to comply with the privacy regulations of the ``right to be forgotten''.

\subsection{Attacks and Defenses for Federated Recommender Systems}
Given the significant advancements achieved by FedRecs, many researchers are now investigating the security concerns associated with these systems.
They have proposed several effective attack methods against FedRecs, exposing the vulnerabilities of current FedRecs under specific conditions.
In general, the attacks in FedRecs can be divided into two categories: inference attacks and poisoning attacks.
Inference attacks aim to detect certain information (e.g., user attributes~\cite{zhang2022comprehensive}, user private date~\cite{yuan2023interaction}) from FedRecs to reveal certain privacy problems, as introduced in Section~\ref{sec_fecrecs}.

In this paper, our topic is closer to poisoning attacks.
According to the attack's goal, there are targeted attacks and untargeted attacks.
Untargeted attacks aim to cause a loss of recommendation accuracy and undermine the validity of the target model.
FedAttack~\cite{wu2022fedattack} attempts to compromise FedRecs using hard negative samples.
Yu et al.~\cite{yu2022untargeted} introduced a cluster-based attack to disrupt recommender systems.
Targeted attacks aim to make specific items to be recommended to as many users as possible.
Compared to untargeted attacks, targeted attacks are more stealthy and are more common due to financial incentives.
Therefore, in this paper, we focus on targeted attacks.
PipAttack~\cite{zhang2022pipattack} demonstrates that malicious users can manipulate the order of item ranks by uploading poisoned gradients.
However, their attack requires a large proportion of compromised clients, which may be unaffordable in real applications.
Rong et al.~\cite{rong2022fedrecattack} reduced the number of malicious users by incorporating more prior knowledge.
~\cite{rong2022poisoning} is the first model poisoning attack that does not rely on any prior knowledge, but its performance is unstable.
Our previous work~\cite{yuan2023manipulating} proposed a more effective model poisoning attack, PSMU, which achieves state-of-the-art performance.
Besides, our previous work is the first to provide a defense mechanism based on gradient clipping to defend against existing model poisoning attacks.

However, all existing targeted poisoning attacks have been launched in FedRecs that rely on collaborative filtering data.
These data suffer from severe sparsity problems, leading to strong biases that make it easier for poisoning attacks to manipulate items, especially for the cold ones.
When visual signals are incorporated, data sparsity can be alleviated. 
Therefore, the performance of existing poisoning attacks for visually-aware FedRecs is unknown.

\subsection{Visually-aware Recommender Systems and Attacks}\label{sec_vrsa_relatedwork}
Visual signals are important for making accurate recommendations~\cite {chen2020try,hung2017computing}.
Some models solely rely on the feature vectors extracted from images to provide recommendations~\cite{jagadeesh2014large,kalantidis2013getting}.
However, the feature vectors are not optimized for making a good recommendation.
VBPR~\cite{he2016vbpr} is the first work that fuses both visual information and collaborative signals based on BPR~\cite{rendle2012bpr}.
After that, many works~\cite{kang2017visually,liu2017deepstyle} are proposed based on the general framework: a pre-trained model is used to extract visual features, and then, certain fusion methods are used to aggregate visual features with collaborative signals (or other modality signals~\cite{yin2015joint,zheng2016keyword,qiu2022contrastive}) to feed a recommendation model.
In this paper, based on such a framework, we extend the basic FedRecs used in our previous work~\cite{yuan2023manipulating} to visually-aware FedRecs.

Due to the large scale of item catalogues, product visual descriptions are usually provided by external sources. 
Since visual information can influence the ranking of items, image providers may have a chance to adversarially manipulate item ranks.
~\cite{di2020taamr} attempts to change the popularity of items with the same categories.
~\cite{liu2021adversarial} is the first work research item promotion via image pollution in centralized recommendation with white-box settings.
~\cite{cohen2021black} further investigates such attacks under black-box settings.
However, all of the above works are based on centralized recommender systems, and they assume that users' recommendation list is available, which is infeasible in FedRecs.
Therefore, in this paper, we explore image poisoning attacks in FedRecs to reveal the threats of incorporating visual information, and then, we propose defense solution to prevent the threat.

\subsection{Diffusion Models}
Motivated by non-equilibrium thermodynamics~\cite{sohl2015deep}, the diffusion model has shown a strong ability to generate high-quality images~\cite{dhariwal2021diffusion,song2021maximum}.
Different from other commonly used generative models such as GANs~\cite{goodfellow2020generative} and VAEs~\cite{kingma2013auto}, diffusion models generate samples by predicting the noise.
Therefore, they are naturally suitable for adversarial image purification~\cite{yang2022diffusion}.
Nie et al.~\cite{nie2022diffusion} was the first to use the diffusion model to purify adversarial images.
Wang et al.~\cite{wang2022guided} utilized guidance to further improve the quality and fidelity of purified images.
This paper takes the first time to integrate diffusion models into the visually-aware federated recommender systems to allow the secure using of visual information.

\section{Visually-aware Federated Recommender Systems}\label{sec_vfedrec}
In this part, we provide the fundamental settings of our visually-aware federated recommender systems.
The federated recommendation framework used in this paper is the same as all previous FedRecs attack works~\cite{zhang2022pipattack,rong2022poisoning,yuan2023interaction}, which was originally proposed by ~\cite{ammad2019federated}.

Let $\mathcal{U}$ and $\mathcal{V}$ denote the set of benign users and items, respectively.
$\left| \mathcal{U}\right|$ and $\left|\mathcal{V}\right|$ are the sizes of users and items.
In FedRec, each user $u_{i}$ is a client who manages its' own training dataset $\mathcal{D}_{i}$.
$\mathcal{D}_{i}$ consists of many user-item interactions $(u_{i}, v_{j}, r_{ij})$, where $r_{ij}$ is a binary rating denoting whether user $u_{i}$ has interacted with item $v_{j}$. 
That is, $r_{ij}=1$ means $u_{i}$ has interacted with $v_{j}$, while $r_{ij}=0$ indicates no interaction between $u_i$ and $v_j$.
In addition, a single image $i_{j}$ is available for each item $v_{j}$ as an auxiliary description, which is uploaded by the item provider and managed by the central server.
$\mathcal{V}_{i}^{+}$ and $\mathcal{V}_{i}^{-}$ are the sets of interacted items and non-interacted items for user $u_{i}$.
Using the above data, FedRec aims to predict $\hat{r}_{ij}$ between user $u_{i}$ and non-interacted item $v_{j}$ and recommend items according to top-K highest prediction scores.

To ensure privacy protection, the parameters of the recommendation model are divided into private and public parameters.
Private parameters are generally user embeddings $\mathbf{U}$, which are maintained by corresponding users and are never shared with others.
The public parameters, on the other hand, include item embeddings $\mathbf{V}$, visual feature extractor $\bm{\Phi}$, visual feature transform matrix $\mathbf{E}$ and other parameters $\bm{\Theta}$, are transmitted between a central server and clients to achieve collaborative learning.

\textbf{Federated learning protocol.}
In FedRecs, a central server coordinates the learning process.
Initially, the central server initializes all public parameters, meanwhile, the clients initialize their corresponding private parameters locally.
Then, a recommender system is trained by iteratively repeating the following steps.
First, the central server randomly selects a set of users $\mathcal{U}_{t-1}$ to participate in the training process and sends the public parameters to these users.
The selected users combine the received public parameters with the private parameters to form a local recommendation model.
The local recommender system is trained on local dataset $\mathcal{D}_{i}$ by optimizing certain objective functions, such as:
\begin{equation}\label{eq_ori_loss}
  \mathcal{L}^{rec} = -\sum\nolimits_{(u_{i}, v_{j}, r_{ij})\in \mathcal{D}_{i}} r_{ij}\log \hat{r}_{ij} + (1-r_{ij})\log (1-\hat{r}_{ij})
\end{equation}
After local training, the selected user $u_{i}$ updates its private parameters (E.q.~\ref{eq_local_update}) and transmits public parameters' gradients $\nabla \mathbf{\Theta}^{t-1}_{i}$, $\nabla \mathbf{E}^{t-1}_{i}$ and $\nabla \mathbf{V}^{t-1}_{i}$ to the central server:
\begin{equation}\label{eq_local_update}
  \mathbf{u}_{i}^{t} = \mathbf{u}_{i}^{t-1} - lr\nabla \mathbf{u}_{i}^{t-1}
\end{equation}
Then, the central server aggregates all received public parameter updates. The following formula takes item embeddings as an example:
\begin{equation}\label{eq_aggregate}
  \begin{aligned}
    &\mathbf{V}^{t} = \mathbf{V}^{t-1} - lr\sum\limits_{u_{i}\in \mathcal{U}_{t-1}} \nabla \mathbf{V}^{t-1}_{i}
  \end{aligned}
\end{equation}
where $lr$ is the learning rate. Note that the visual extractor $\mathbf{\Phi}$ can be trainable or freeze. In this paper, we freeze the visual extractor $\mathbf{\Phi}$, since it is a large pretrained model and retraining it will dramatically increase the difficulty of convergence.

\textbf{Base recommendation model.}
Generally, the above federated recommendation framework is compatible with most existing deep learning-based recommendation models~\cite{zhang2019deep}.
In our previous work~\cite{yuan2023manipulating}, we choose two classical and widely used recommenders, Neural Collaborative Filtering (NCF)~\cite{he2017neural} and LightGCN~\cite{he2020lightgcn}, as the base model.
In this paper, we extend these two models to consider visual information when making recommendations, namely VNCF and LightVGCN.
And then, we integrate these two models into the above FedRec framework to form Fed-VNCF and Fed-LightVGCN.

In Fed-VNCF, the local recommendation model in client $u_{i}$ predicts $\hat{r}_{ij}$ using the following formula:
\begin{equation}\label{eq_ncf}
  \hat{r}_{ij} = \sigma (\mathbf{h}^\top FFN([\mathbf{u}_{i}, \mathbf{v}_{j}, \mathbf{E}\bm{\Phi}(i_{j})]))
\end{equation}
where $\mathbf{h}$ and $\mathbf{E}$ are trainable public parameters, $\mathbf{u}_{i}$ and $\mathbf{v}_{j}$ are embeddings of user $u_{i}$ and item $v_{j}$, and $[\cdot]$ is concatenation operation.

For Fed-LightVGCN, the user-item interactions are viewed as a bipartite graph and all users and items are treated as distinct nodes.
Then, user and item embeddings are learned by propagating their neighbour nodes' embeddings:
\begin{equation}
  \begin{aligned}\label{eq_lightgcn}
    &\mathbf{u}_{i}^{l} = \sum\limits_{j\in \mathcal{N}_{u_{i}}}\frac{1}{\sqrt{\left| \mathcal{N}_{u_{i}} \right|} \sqrt{\left| \mathcal{N}_{v_{j}} \right|}}(\mathbf{v}_{j}^{l-1} + \mathbf{E}\bm{\Phi}(i_{j}))\\
    &\mathbf{v}_{j}^{l} = \sum\limits_{i\in \mathcal{N}_{v_{j}}}\frac{1}{\sqrt{\left| \mathcal{N}_{v_{j}} \right|} \sqrt{\left| \mathcal{N}_{u_{i}} \right|}}\mathbf{u}_{i}^{l-1}
  \end{aligned}
\end{equation}
where $\mathcal{N}_{u_{i}}$ and $\mathcal{N}_{v_{j}}$ denote the sets of $u_i$'s and $v_j$'s neighbors.
$l$ is the propagation layer.
In order to protect privacy, each user can only perform the above calculation on its local bipartite graph.
After propagation, we aggregate all layers' embeddings as the final user and item embeddings:
\begin{equation}
  \mathbf{u}_{i} = \sum\limits_{l=0}^{L} \mathbf{u}_{i}^{l}, \quad \mathbf{v}_{j} = \sum\limits_{l=0}^{L} \mathbf{v}_{j}^{l} 
\end{equation}
Finally, the same as VNCF, we use E.q.~\ref{eq_ncf} to compute the predicted preference scores.

\begin{algorithm}[!ht]
  \renewcommand{\algorithmicrequire}{\textbf{Input:}}
  \renewcommand{\algorithmicensure}{\textbf{Output:}}
  \caption{Visually-aware Federated Recommender Systems.} \label{alg_fedrec}
  \begin{algorithmic}[1]
    \Require global epoch $T$; local epoch $L$; learning rate $lr$, visual extractor $\mathbf{\Phi}$ \dots
    \Ensure public parameter $\mathbf{V}$, $\mathbf{E}$ and $\mathbf{\Theta}$, local client embedding $\mathbf{u}_{i}|_{i \in \mathcal{U}}$
    \State Initialize public parameter $\mathbf{V}^{0}$, $\mathbf{E}^{0}$, and $\mathbf{\Theta}^{0}$
    \State Initialize item image set $\mathcal{I}^{0}=\{\}$
    \For {each round t =1, ..., $T$}
    \If{new images uploaded by item providers}
      \State $\mathcal{I}^{t}\leftarrow$ update $\mathcal{I}^{t-1}$ // The threats of image poisoning attack
    \EndIf
      \State sample a fraction of clients $\mathcal{U}_{t-1}$ from $\mathcal{U}$  // The threats of model poisoning attack
        \For{$u_{i}\in \mathcal{U}_{t-1}$ \textbf{in parallel}}
          \State // run on client $u_{i}$
          \State calculate $\nabla \mathbf{u}_{i}^{t-1}$, $\nabla \mathbf{V}_{i}^{t-1}$, and $\nabla \mathbf{E}^{t-1}_{i}$, and $\nabla \mathbf{\Theta}_{i}^{t-1}$ using E.q.~\ref{eq_ori_loss}
          \State $\mathbf{u}_{i}^{t}\leftarrow$ update local private parameters using E.q.~\ref{eq_local_update}
          \State upload $\nabla \mathbf{V}_{i}^{t-1}$, $\nabla \mathbf{E}_{i}^{t-1}$ and  $\nabla \mathbf{\Theta}_{i}^{t-1}$ to the central server
        \EndFor
      \State $\mathbf{V}^{t}, \mathbf{E}^{t}, \mathbf{\Theta}^{t}\leftarrow$ aggregate gradients using E.q.~\ref{eq_aggregate}
    \EndFor
    \end{algorithmic}
\end{algorithm}


\section{Adversarial Item Promotion via Image and Model Poisoning Attacks}\label{sec_attack}
In this section, we present the details of our attacks, including the model poisoning attack (PSMU), image poisoning attack (PSMU(V)), and the combination of model poisoning attack and image poisoning attack (PSMU++).

\subsection{Attack Task Formulation}
Manipulating recommender systems includes promoting and demoting the rank order of items.
In this work, we mainly discuss item promotion, since the demotion can be achieved by reversing the attack objective or promoting all other items.
Adversarial item promotion has been widely studied in the poisoning attacks for FedRecs, which aims to increase the target item's exposure chances motivated by financial incentives~\cite{zhang2022pipattack}.
However, all previous works only consider item promotion in FedRecs with collaborative data, none of them investigates the threat in visually-aware FedRecs.
In this section, we present preliminaries and basic settings of adversarial item promotion in this paper.

\textbf{Attack goal.}
Obviously, the goal of adversaries is to promote target items to as many users as possible.
Formally, given that a recommender system recommends $K$ items $\hat{\mathcal{V}}_{i}$ to user $u_{i}$, the adversaries would like to improve the target item's exposure rate at rank $K$ (ER@K):
\begin{equation}
  \label{eq_er}
  ER@K = \frac{1}{\left|\widetilde{\mathcal{V}} \right|} \sum\limits_{v_{j} \in \widetilde{\mathcal{V}}} \frac{\left|\left\{ u_{i} \in \mathcal{U} | v_{j}\in \hat{\mathcal{V}}_{i} \land v_{j}\in \mathcal{V}_{i}^{-} \right\}\right|}{\left|\left\{ u_{i} \in \mathcal{U} | v_{j}\in \mathcal{V}_{i}^{-} \right\} \right|} 
\end{equation}
$\widetilde{\mathcal{V}}$ is the set of target items.

\textbf{Attack approach.}
We explore two kinds of poisoning attacks in this work: model poisoning attacks (a.k.a., gradient poisoning attacks) and image poisoning attacks.
For model poisoning attacks, as shown in the upper part of Fig.~\ref{fig_image_poisoning}, the attacker will employ a group of malicious users $\widetilde{\mathcal{U}}$ to upload poisoned gradients to optimize E.q.~\ref{eq_er}:
\begin{equation}
  \label{eq_simple_opt}
  \begin{aligned}
    \mathop{argmax}\limits_{\{\nabla \widetilde{\mathbf{V}}^{t}, \nabla \widetilde{\mathbf{E}}^{t}, \nabla \widetilde{\mathbf{\Theta}}^{t}\}_{t=s}^{T-1}} ER@K(\mathbf{U}^{T},\mathbf{V}^{T},\mathbf{E}^{T},\mathbf{\Theta}^{T})\\
  \end{aligned}
\end{equation}
where $s$ is the epoch when the attacks are launched. $\nabla \widetilde{\mathbf{V}}^{t}$, $\nabla \widetilde{\mathbf{E}}^{t}$ and $\nabla \widetilde{\mathbf{\Theta}}^{t}$ are gradients generated by malicious users at epoch $t$.
These poisoned gradients will be aggregated in the central server using E.q.~\ref{eq_aggregate}, since the central server is unaware of poisoned attacks.

For image poisoning attacks, we assume that the adversary is the target item image provider. 
It increases the target item's ER@K by uploading an image with human-unaware perturbations as follows:
\begin{equation}
  \label{eq_img_opt}
  \begin{aligned}
    &\mathop{argmax}\limits_{\{\widetilde{\mathcal{I}}^{t}\}_{t=s}^{T-1}} ER@K(\mathbf{U}^{T},\mathbf{V}^{T}, \mathbf{E}^{T}, \mathbf{\Theta}^{T})\\
    &\widetilde{\mathbf{i}}_{j}^{t} = \mathbf{i}_{j} + \boldsymbol{\delta}_{j}^{t},\, \text{for } \widetilde{\mathbf{i}}_{j}^{t} \in \widetilde{\mathcal{I}}^{t}
  \end{aligned}
\end{equation}
where $\widetilde{\mathcal{I}}^{t}$ is the set of poisoned images for target items $\widetilde{\mathcal{V}}$ at epoch $t$.
Note that the model poisoning attacks and image poisoning attacks can be launched simultaneously to promote the same target items, as follows:
\begin{equation}
  \label{eq_img_and_mp_opt}
  \begin{aligned}
    &\mathop{argmax}\limits_{\{\nabla \widetilde{\mathbf{V}}^{t}, \nabla \widetilde{\mathbf{E}}^{t}, \nabla \widetilde{\mathbf{\Theta}}^{t},\widetilde{\mathcal{I}}^{t}\}_{t=s}^{T-1}} ER@K(\mathbf{U}^{T},\mathbf{V}^{T}, \mathbf{E}^{T}, \mathbf{\Theta}^{T})\\
  \end{aligned}
\end{equation}

\textbf{Attack prior knowledge.}
Following our previous work~\cite{yuan2023manipulating}, in this paper, we assume that for both model poisoning attacks and image poisoning attacks, the attacker knows public parameters $\mathbf{V}$, $\mathbf{E}$, and $\mathbf{\Theta}$ received from the central server, which is consistent with FedRecs' protocol.
Besides, we assume the image poisoning attacker already knows the visual extractor $\mathbf{\Phi}$, which is reasonable since $\mathbf{\Phi}$ is an open-source pretrained model and it can be easily inferred by comparing the image feature vectors generated from the system and from guessed extractors.

\subsection{Poisoning Attack}
\textbf{Formulate the optimization problem.} The goal of all our attacks, including PSMU, PSMU(V), and PSMU++, is to promote target items $\widetilde{\mathcal{V}}$ to as many users as possible.
To achieve that, these attacks use different approaches to maximize E.q.~\ref{eq_er}, such as E.q.~\ref{eq_simple_opt}, E.q.~\ref{eq_img_opt}, and E.q.~\ref{eq_img_and_mp_opt}.
However, for all these attacks, it is challenging to directly optimize their objective function because of the following two problems:
(1) The complex dependence of model parameter updates~\cite{yuan2023manipulating}; (2) ER@K is not differentiable; 
For the first problem, instead of finding a globally optimal solution, we greedily calculate the optimal results at each epoch, which will simplify the optimization problem:
\begin{equation}
  \label{eq_iter_er}
  \begin{aligned}
    \text{PSMU:}\quad &\mathop{argmax}\limits_{\{\nabla \widetilde{\mathbf{V}}^{t-1}, \nabla \widetilde{\mathbf{\Theta}}^{t-1}\}} ER@K(\mathbf{U}^{t-1},\mathbf{V}^{t-1} - lr\nabla \widetilde{\mathbf{V}}^{t-1}, \mathbf{E}^{t-1} - lr\nabla \widetilde{\mathbf{E}}^{t-1} ,\mathbf{\Theta}^{t-1} - lr\nabla \widetilde{\mathbf{\Theta}}^{t-1})\\
    \text{PSMU(V):}\quad &\mathop{argmax}\limits_{\{\widetilde{\mathcal{I}}^{t-1}\}} ER@K(\mathbf{U}^{t-1},\mathbf{V}^{t-1}, \mathbf{E}^{t-1},\mathbf{\Theta}^{t-1})\\
    \text{PSMU++:}\quad &\mathop{argmax}\limits_{\{\nabla \widetilde{\mathbf{V}}^{t-1}, \nabla \widetilde{\mathbf{\Theta}}^{t-1},\widetilde{\mathcal{I}}^{t-1}\}} ER@K(\mathbf{U}^{t-1},\mathbf{V}^{t-1} - lr\nabla \widetilde{\mathbf{V}}^{t-1},\mathbf{E}^{t-1} - lr\nabla \widetilde{\mathbf{E}}^{t-1},\mathbf{\Theta}^{t-1} - lr\nabla \widetilde{\mathbf{\Theta}}^{t-1})\\
  \end{aligned}
\end{equation}
For the second problem, we approximately optimize ER@K by forcing the target items' predicted preference scores to be higher than other recommended items' scores:
\begin{equation}
  \label{eq_naive_loss}
  \mathcal{L}^{att} = \sum\limits_{u_{i}\in \mathcal{U}} \sum\limits_{v_{t} \in \widetilde{\mathcal{V}} \land v_{t} \notin \mathcal{V}_{i}^{+}} \sum\limits_{v_{j} \in \hat{\mathcal{V}_{i}} \land v_{j} \notin \widetilde{\mathcal{V}}} \sigma (\hat{r}_{ij} - \hat{r}_{it})
\end{equation}
We omit the time index for the prediction scores in E.q.~\ref{eq_naive_loss} to make the formula clear.
To compute E.q.~\ref{eq_naive_loss}, we need the feedback from benign users (i.e., the recommended items $\hat{\mathcal{V}_{i}}$ and the scores of target items $\hat{r}_{it}$).
The calculation of $\hat{\mathcal{V}_{i}}$ and $\hat{r}_{it}$  are based on user $u_{i}$'s private parameters, which is strictly not accessible in FedRecs.
Following our previous work~\cite{yuan2023manipulating}, we construct a group of synthetic users to replace benign users.
Specifically, we randomly select a group of items as fake user $\widetilde{u}_{i}$'s interacted items $\widetilde{\mathcal{V}}_{i}^{+}$.
Based on $\widetilde{\mathcal{V}}_{i}^{+}$, we build the synthetic dataset $\widetilde{\mathcal{D}}_{i}$.
Then, we fix all public parameters and train the fake user's user embeddings with the recommendation objective:
\begin{equation}
  \label{eq_avid_user}
  \widetilde{\mathbf{U}}^{t-1} = \mathop{argmin}\limits_{\widetilde{\mathbf{U}}^{t-1}} \mathcal{L}^{rec}(\widetilde{\mathbf{U}}^{t-1}, \mathbf{V}^{t-1}, \mathbf{E}^{t-1},\mathbf{\Theta}^{t-1}, \widetilde{\mathcal{D}}^{t-1})
\end{equation}
Note that at different epochs, we reconstruct different $\widetilde{\mathcal{V}}_{i}^{+}$ and $\widetilde{\mathcal{D}}_{i}$, so that even with a small size of malicious users, we can still simulate many synthetic users.

After constructed synthetic users, E.q.~\ref{eq_naive_loss} is transformed to:
\begin{equation}
  \label{eq_avid_loss}
  \widetilde{\mathcal{L}}^{att} = \sum\limits_{\widetilde{u}_{i}\in \widetilde{\mathcal{U}}} \sum\limits_{v_{t} \in \widetilde{\mathcal{V}} \land v_{t} \notin \widetilde{\mathcal{V}}^{+}_{i}} \sum\limits_{v_{j} \in \hat{\widetilde{\mathcal{V}}_{i}} \land v_{j} \notin \widetilde{\mathcal{V}}} \sigma (\hat{r}_{ij} - \hat{r}_{it})
\end{equation}
where $\hat{\widetilde{\mathcal{V}}_{i}}$ is the set of items that have the highest prediction scores for malicious user $\widetilde{u}_{i}$.

\textbf{Optimizing via poisoned gradients (PSMU).}
As shown in E.q.~\ref{eq_iter_er}, PSMU promotes target items by uploading poisoned gradients to the central server. 
Specifically, given current public parameters $\mathbf{V}^{t-1}$, $\mathbf{E}^{t-1}$, $\mathbf{\Theta}^{t-1}$, and the calculated fake user embeddings $\widetilde{\mathbf{U}}^{t-1}$, we compute the poisoned gradients as follows:
\begin{equation}
  \label{eq_model_poison_grad}
  \begin{aligned}
    &\nabla \widetilde{\mathbf{V}}^{t-1} = \frac{\partial}{\partial \mathbf{V}^{t-1}} \widetilde{\mathcal{L}}^{att}(\widetilde{\mathbf{U}}^{t-1},\mathbf{V}^{t-1},\mathbf{E}^{t-1},\mathbf{\Theta}^{t-1})  \\
    &\nabla \widetilde{\mathbf{E}}^{t-1} = \frac{\partial}{\partial \mathbf{E}^{t-1}} \widetilde{\mathcal{L}}^{att}(\widetilde{\mathbf{U}}^{t-1},\mathbf{V}^{t-1},\mathbf{E}^{t-1}, \mathbf{\Theta}^{t-1})  \\
    &\nabla \widetilde{\mathbf{\Theta}}^{t-1} = \frac{\partial}{\partial \mathbf{\Theta}^{t-1}} \widetilde{\mathcal{L}}^{att}(\widetilde{\mathbf{U}}^{t-1},\mathbf{V}^{t-1},\mathbf{E}^{t-1}, \mathbf{\Theta}^{t-1})  \\
  \end{aligned}
\end{equation} 
It is worth noting that compared to original poison FedRecs, in visually-aware FedRecs, we also poison public parameters related to visual signals, such as $\mathbf{E}$, for a fair comparison.
Naturally, poisoning more items will make the attack goal easier to achieve, however, it will also cause too many side effects on FedRec performance.
Therefore, for item embeddings, PSMU only uploads poisoned gradients for target items' embeddings.
\begin{equation}
  \label{eq_refine}
  \nabla \widetilde{\mathbf{V}}^{t-1}=
  \begin{cases}
  \mathbf{0}& v_{m} \notin \widetilde{V}\\
  \nabla \widetilde{\mathbf{V}}^{t-1}_{m}& v_{m} \in \widetilde{V}
  \end{cases} \quad m=0,1,\dots,\left| \mathcal{V} \right|
\end{equation}

\textbf{Optimizing via poisoned images (PSMU(V)).}
In visually-aware FedRecs, item images are provided by external sources, which are usually item providers. 
These item providers may provide images with slight pollutions to mislead recommender systems to give higher preference scores to their items for as many users as possible.
Formally, given $\mathbf{V}^{t-1}$, $\mathbf{E}^{t-1}$, $\mathbf{\Theta}^{t-1}$ and $\widetilde{\mathbf{U}}^{t-1}$, the attacker calculates perturbations as follows:
\begin{equation}
  \label{eq_image_poison_grad}
  \begin{aligned}
    &\bm{\delta}^{t-1} = \mathop{argmin}\limits_{\bm{\delta}^{t-1}} \widetilde{\mathcal{L}}^{att}(\widetilde{\mathbf{U}}^{t-1},\mathbf{V}^{t-1},\mathbf{E}^{t-1},\mathbf{\Theta}^{t-1}), \quad \left\| \bm{\delta} \right\| \le \epsilon\\
  \end{aligned}
\end{equation}
To avoid the perturbations being aware by normal users, the attacker should restrict the size of noise $\bm{\delta}^{t-1}$ with the bound $\epsilon$ at each epoch.
Previous visual attacks in centralized recommender systems usually set $\epsilon$ to at least $32$ for a $255$ pixel value range. In this paper, we only use $\epsilon=4$ then we can achieve ER@5=1.0 results, which shows the severe threats of image poisoning attacks in FedRecs.

\textbf{PSMU++.}
Since PSMU and PSMU(V) have consistent objectives and there is no conflict between their poisoning implementations, the adversary can combine these two attacks together to form a more effective item promotion attack, named PSMU++.
In PSMU++, we assume the item providers are the adversaries and they not only upload poisoned product images, but also upload poisoned gradients using a group of compromised users.
Specifically, the malicious users upload poisoned gradients by launching the PSMU algorithm, meanwhile, the item provider launch PSMU(V) based on these malicious users' synthetic user embeddings (i.e., Algorithm~\ref{alg_psmuv} Line 4).

\begin{algorithm}[!ht]
  \renewcommand{\algorithmicrequire}{\textbf{Input:}}
  \renewcommand{\algorithmicensure}{\textbf{Output:}}
  \caption{PSMU: Poisoning with Synthetic Malicious Users} \label{alg_psmu}
  \begin{algorithmic}[1]
     
    \Require public parameters $\mathbf{V}^{t-1}$, $\mathbf{\Theta}^{t-1}$, $\mathbf{E}$
    \Ensure public parameter poisoned gradients $\nabla \widetilde{\mathbf{V}}_{i}^{t-1}$, $\nabla \widetilde{\mathbf{E}}_{i}^{t-1}$, $\nabla \widetilde{\mathbf{\Theta}}_{i}^{t-1}$
    \State // run on malicious client $\widetilde{u}_{i}$
    \State randomly construct training set $\widetilde{\mathcal{D}}^{t-1}_{i}$
    \State calculate synthetic user embedding $\widetilde{\mathbf{u}}_{i}^{t-1}$ using E.q.~\ref{eq_avid_user}
    \State calculate $\nabla \widetilde{\mathbf{V}}_{i}^{t-1}$, $\nabla \widetilde{\mathbf{E}}_{i}^{t-1}$, $\nabla \widetilde{\mathbf{\Theta}}_{i}^{t-1}$ using E.q.~\ref{eq_model_poison_grad}
    \State $\nabla \widetilde{\mathbf{V}}_{i}^{t-1}\leftarrow$ constraint $\nabla \widetilde{\mathbf{V}}_{i}^{t-1}$ using E.q.~\ref{eq_refine}
    \State upload $\nabla \widetilde{\mathbf{V}}_{i}^{t-1}$,  $\nabla \widetilde{\mathbf{E}}_{i}^{t-1}$, $\nabla \widetilde{\mathbf{\Theta}}_{i}^{t-1}$ to the central server
    \end{algorithmic}
\end{algorithm}

\begin{algorithm}[!ht]
  \renewcommand{\algorithmicrequire}{\textbf{Input:}}
  \renewcommand{\algorithmicensure}{\textbf{Output:}}
  \caption{PSMU(V): Poisoning with Synthetic Malicious Users via Visual Information} \label{alg_psmuv}
  \begin{algorithmic}[1]
     
    \Require public parameters $\mathbf{V}^{t-1}$, $\mathbf{\Theta}^{t-1}$, $\mathbf{E}$
    \Ensure image adversarial perturbation $\bm{\delta}^{t-1}$
    
    \State // run on target items' image provider
    \If {malicious users $\widetilde{\mathcal{U}}$ exist}
      \State // PSMU++ 
      \State request for malicious user embeddings $\widetilde{\mathbf{U}}^{t-1}$
    \Else
      \State randomly construct training set $\widetilde{\mathcal{D}}^{t-1}$ and calculate user embeddings $\widetilde{\mathbf{U}}^{t-1}$ using E.q.~\ref{eq_avid_user}
    \EndIf
    
    \State calculate $\bm{\delta}^{t-1}$ using E.q.~\ref{eq_image_poison_grad}
    \State construct and upload adversarial images $\widetilde{\mathcal{I}}^{t-1}$
    \end{algorithmic}
\end{algorithm}

\begin{figure}[!htbp]
  \centering
  \includegraphics[width=\linewidth]{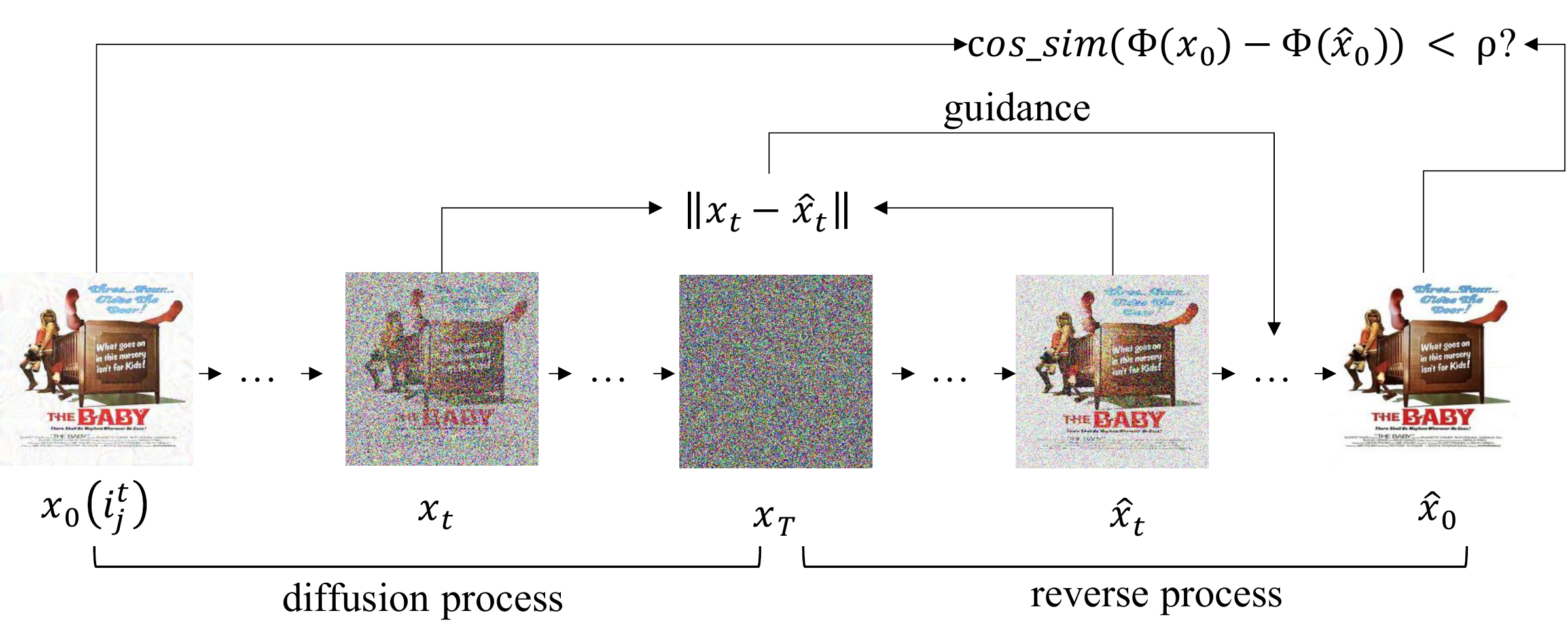}
  \caption{Overview of the proposed image poisoning defender: GDMPD.}\label{fig_gdmpf_overview}
  \Description{Overview of image poisoning attack.}
\end{figure}
\section{Guided Diffusion Model for Purification and Detection}\label{sec_defense}
The experimental results in Section~\ref{sec_exp} indicate that traditional model poisoning attacks, such as PSMU, are ineffective in visually-aware FedRecs. However, the presence of visual information creates another backdoor, which provides an opportunity for corporations to promote items effectively through PSMU and PSMU(V), underscoring the urgent need for image poisoning defense. 
GDMPD leverages a pretrained Denoising Diffusion Probabilistic Model (DDPM)~\cite{sohl2015deep,ho2020denoising} to purify all uploaded images with guidance, eliminating the need for additional computation resources for training.
After purification, adversarial images have a high probability of losing their delicate perturbations. 
To maintain the recommender system, it is essential to detect which images are adversarial. Based on the purified image, our GDMPD achieves an adversarial image detection function. Notably, our detection method is also training-free. 
Fig.~\ref{fig_gdmpf_overview} gives an overview illustration of our GDMPD.

\subsection{Diffused Image Guided Purification}
In Section~\ref{sec_attack}, we use $\mathbf{i}_{j}^{t}$ to represent the visual discription of item $j$ uploaded at epoch $t$.
For clarity of description, when using the image as input for GDMPD, we denote it as $\mathbf{x}$.
Generally, DDPM consists of two Markov processes: the diffusion process and the reverse process.
In the diffusion process, DDPM adds noise to the input image at each time step until it becomes Gaussian noise.
Then, the reverse process gradually removes this noise to recover the input image.

\textbf{Diffusion process.} Formally, assume $\mathbf{x}_{0}$ be the input image where $t=0$. Note that to avoid misunderstanding with FedRec's global epochs $t$ denoted by superscript, here we use subscript $t$ to denote the diffusion time step.
$T$ is the length of diffusion steps.
DDPM incrementally corrupts the input image $\mathbf{x}_{0}$ into Gaussian noise as follows:
\begin{equation}
  \label{eq_ddpm_diffusion}
  \begin{aligned}
    &q(\mathbf{x}_{1}, \mathbf{x}_{2}, \dots, \mathbf{x}_{T}|\mathbf{x}_{0}) = \prod\limits_{t=1}^{T} q(\mathbf{x}_{t}|\mathbf{x}_{t-1}) \\
    &q(\mathbf{x}_{t}|\mathbf{x}_{t-1})=\mathcal{N}(\mathbf{x}_{t}; \sqrt{1-\beta_{t}}\mathbf{x}_{t-1},\beta_{t}\mathbf{I})
  \end{aligned}
\end{equation}
where $\mathcal{N}(x, \mu, \sigma^{2})$ means $x$ is sampled from a Gaussian distribution with a mean $\mu$ and variance $\sigma$.
$\beta_{t}$ is generated from a predefined noise adding schedule $\beta$. 
Common settings of $\beta$ include consine~\cite{ho2020denoising}, square-root~\cite{li2022diffusion}, and linear schedule~\cite{wang2022guided}.
In this paper, following~\cite{wang2022guided,dhariwal2021diffusion}, we use the linear schedule and define $\beta_{t}$ as: $\beta_{t} = \frac{t-1}{T-1}(\beta_{T} - \beta_{1})$, where $\beta_{T}=2\times 10^{-2}$ and $\beta_{1}=1\times 10^{-4}$ are hyper-parameters.
According to~\cite{ho2020denoising}, we can directly calculate $\mathbf{x}_{t}$ at an arbitrary diffusion step directly conditioned on $\mathbf{x}_{0}$ with following euqation:
\begin{equation}
  \label{eq_ddpm_diffusion_alpha}
  \begin{aligned}
    &q(\mathbf{x}_{t}|\mathbf{x}_{0})=\mathcal{N}(\mathbf{x}_{t}; \sqrt{\bar{\alpha}_{t}}\mathbf{x}_{0},(1-\bar{\alpha}_{t})\mathbf{I})\\
    &\bar{\alpha}_{t} = \prod\limits_{i=1}^{t}\alpha_{i},\quad \alpha_{i} = 1 - \beta_{i}
  \end{aligned}
\end{equation}
Then, with reparameter trick, we can generate $\mathbf{x}_{t}$ as follows:
\begin{equation}
  \label{eq_ddpm_reparameter}
  \begin{aligned}
    &\mathbf{x}_{t} = \sqrt{\bar{\alpha}_{t}}\mathbf{x}_{0} + \sqrt{1-\bar{\alpha}_{t}}\bm{\zeta}
  \end{aligned}
\end{equation}
where $\bm{\zeta}$ is noise sampled from standard Gaussian distribution, i.e., $\bm{\zeta}\sim \mathcal{N}(\mathbf{0}, \mathbf{I})$.

Considering the input image is adversarial image $\mathbf{x}_{0}^{adv}$ (a.k.a., $\widetilde{\mathbf{i}}$ in Section~\ref{sec_attack}) and $\mathbf{x}_{0}^{adv}=\mathbf{x}_{0} + \bm{\delta}$, then, after $t$ steps diffusion, the image equals to:
\begin{equation}
  \label{eq_ddpm_adv_diffuse}
  \begin{aligned}
    &\mathbf{x}^{adv}_{t} = \sqrt{\bar{\alpha}_{t}}\mathbf{x}_{0} + \sqrt{\bar{\alpha}_{t}}\bm{\delta} + \sqrt{1-\bar{\alpha}_{t}}\bm{\zeta}
  \end{aligned}
\end{equation}
When $t$ increases, $\sqrt{\bar{\alpha}_{t}}$ will gradually decrease while $\sqrt{1-\bar{\alpha}_{t}}$ will gradually increase.
Since $\left\| \bm{\delta} \right\|$ should be small (lower than $\epsilon$) to guarantee the perturbations' unawareness, after an appropriate length of diffusion, the magnitude of Gaussian noise $\bm{\zeta}$ will be large enough to submerge the delicately calculated perturbation $\bm{\delta}$.
Meanwhile, the semantic meaning of original image $\mathbf{x}_{0}$ still can be largely preserved, since $\bm{\delta}$ is negligible compared with $\mathbf{x}_{0}$.

\textbf{Reverse process.} The reverse process is a Markov process that denoises the diffused $\mathbf{x}_{t}$ to approximate the original input $\mathbf{x}_{0}$ by predicting the noise added in the diffusion process.
Formally, the reverse process from step $T$ to $0$ is as follows:
\begin{equation}
  \label{eq_ddpm_reverse}
  \begin{aligned}
    &p_{\mathbf{w}}(\hat{\mathbf{x}}_{0}, \hat{\mathbf{x}}_{1}, \dots, \hat{\mathbf{x}}_{T-1}|\mathbf{x}_{T}) = \prod\limits_{t=1}^{T} p_{\mathbf{w}}(\hat{\mathbf{x}}_{t-1}|\hat{\mathbf{x}}_{t}) \\
    &p_{\mathbf{w}}(\hat{\mathbf{x}}_{t-1}|\hat{\mathbf{x}}_{t}) = \mathcal{N}(\hat{\mathbf{x}}_{t-1};\bm{\mu_{\mathbf{w}}}(\hat{\mathbf{x}}_{t},t), \Sigma_{\mathbf{w}}(\hat{\mathbf{x}}_{t},t)\mathbf{I})
  \end{aligned}
\end{equation}
where the mean $\bm{\mu_{\mathbf{w}}}(\mathbf{x}_{t},t)$ is a neural network parameterized by $\mathbf{w}$, the variance $\Sigma_{\mathbf{w}}(\mathbf{x}_{t},t)$ can be either neural network or predefined time step dependent constant~\cite{ho2020denoising,nichol2021improved}.
As results, the reverse process iteratively samples $\hat{\mathbf{x}}_{t-1}$ using $p_{\mathbf{w}}(\hat{\mathbf{x}}_{t-1}|\hat{\mathbf{x}}_{t})$ to get the predicted input image $\hat{\mathbf{x}}_{0}$.
Assume the input image is adversarial image $\mathbf{x}_{0}^{adv}$, after the process of E.q.~\ref{eq_ddpm_adv_diffuse}, the adversarial perturbations are corrupt by gradually added Gaussian noise.
Then, E.q.~\ref{eq_ddpm_reverse} is used to eliminate the Gaussian noise and is very likely to simultaneously remove the perturbations.
This is because: 
(1) The normalization of perturbation is small and the adversarial information is destroyed by adding noise in $\mathbf{x}^{adv}_{t}$;
(2) We use a pretrained DDPM which is learned on normal image datasets, therefore, it tends to recover the image to the domain of clean images in the reverse process. 

\textbf{Guided reverse process.} 
However, simply using DDPM to purify uploaded images in FedRecs will have the following challenge: how to largely recover the original semantic of the input image meanwhile remove most perturbations.
Specifically, if the diffusion steps $t$ are too large, the original information of $\mathbf{x}_{0}$ in $\mathbf{x}_{t}$ will be destroyed and the results of the reverse process will tend to be random~\cite{dhariwal2021diffusion}.
As a result, the FedRecs' recommendation performance will be compromised since all items' visual information will be altered by the defense mechanism.
On the contrary, when the diffusion steps are too small, the diffusion and reverse process may not be strong enough to purify all perturbations.
In FedRecs, since perturbations are usually small, when we diffuse the image to Gaussian distribution, the perturbations will be largely be submerged.
Therefore, the challenge is mainly about how to recover an image with high quality.

To improve the fidelity of images generated by the reverse process, we propose to add guidance during the reverse process.
Concretely, we use the counterpart image $\mathbf{x}_{t}$ in the diffusion steps to guide $\hat{\mathbf{x}}_{t}$'s generation as shown in Fig.~\ref{fig_gdmpf_overview}.
To achieve this, we modify the reverse process $p_{\mathbf{w}}(\hat{\mathbf{x}}_{t-1}|\hat{\mathbf{x}}_{t})$ in E.q.~\ref{eq_ddpm_reverse} to condition on $\mathbf{x}_{t}$, i.e., $p_{\mathbf{w}}(\hat{\mathbf{x}}_{t-1}|\hat{\mathbf{x}}_{t}, \mathbf{x}_{t})$.
According to~\cite{sohl2015deep,dhariwal2021diffusion}, we can further get the following approximation: 
\begin{equation}
  \label{eq_ddpm_reverse_guide1}
  \begin{split}
   \log p_{\mathbf{w}}(\hat{\mathbf{x}}_{t-1}|\hat{\mathbf{x}}_{t}, \mathbf{x}_{t}) &\approx \log p_{\mathbf{w}}(\hat{\mathbf{x}}_{t-1}|\hat{\mathbf{x}}_{t})p(\mathbf{x}_{t}|\hat{\mathbf{x}}_{t}) \\
   &\approx \log p(z)\\
  \end{split}
\end{equation}
\begin{equation}
  \label{eq_ddpm_reverse_z}
  \begin{aligned}
   z \sim \mathcal{N} (\bm{\mu_{\mathbf{w}}}(\hat{\mathbf{x}}_{t},t) + \Sigma_{\mathbf{w}}(\hat{\mathbf{x}}_{t},t) \nabla_{\hat{\mathbf{x}}_{t}}\log p(\mathbf{x}_{t}|\hat{\mathbf{x}}_{t}), \Sigma_{\mathbf{w}}(\hat{\mathbf{x}}_{t},t)\mathbf{I} )
  \end{aligned}
\end{equation}
where $p_{\mathbf{w}}(\hat{\mathbf{x}}_{t-1}|\hat{\mathbf{x}}_{t})$ is the probability from unconditional DDPM and $p(\mathbf{x}_{t}|\hat{\mathbf{x}}_{t})$ can be interpreted as ``how close $\mathbf{x}_{t}$ and $\hat{\mathbf{x}}_{t}$ are''.
In this paper, $p(\mathbf{x}_{t}|\hat{\mathbf{x}}_{t})$ is designed as follows:
\begin{equation}
  \label{eq_ddpm_guidance_desgin}
  \begin{aligned}
    p(\mathbf{x}_{t}|\hat{\mathbf{x}}_{t}) = \text{exp}(\lambda\left\|\mathbf{x}_{t} - \hat{\mathbf{x}}_{t} \right\|)
  \end{aligned}
\end{equation}
$\lambda$ is the factor that controls the scale of guidance. $\left\|\cdot\right\|$ is mean squared error. Combined with E.q.~\ref{eq_ddpm_reverse_z}, we can get:
\begin{equation}
  \label{eq_ddpm_guidance_new_z}
  \begin{aligned}
    z \sim \mathcal{N} (\bm{\mu_{\mathbf{w}}}(\hat{\mathbf{x}}_{t},t) + \lambda\Sigma_{\mathbf{w}}(\hat{\mathbf{x}}_{t},t) \nabla_{\hat{\mathbf{x}}_{t}}\left\|\mathbf{x}_{t} - \hat{\mathbf{x}}_{t} \right\|, \Sigma_{\mathbf{w}}(\hat{\mathbf{x}}_{t},t)\mathbf{I} )
  \end{aligned}
\end{equation}
Finally, we can use the pretrained DDPM to infer purified image $\hat{\mathbf{x}}_{0}$ given the input image $\mathbf{x}_{0}$.
Algorithm~\ref{alg_ddpm_guidance} illustrates how to purify uploaded item visual information using our diffused image-guided DDPM.

\begin{algorithm}[!ht]
  \renewcommand{\algorithmicrequire}{\textbf{Input:}}
  \renewcommand{\algorithmicensure}{\textbf{Output:}}
  \caption{Diffused Image Guided Purification} \label{alg_ddpm_guidance}
  \begin{algorithmic}[1]
     
    \Require pretrained DDPM $\bm{\mu_{\mathbf{w}}}(\hat{\mathbf{x}}_{t},t)$ and $\Sigma_{\mathbf{w}}(\hat{\mathbf{x}}_{t},t)$, guidance factor $\lambda$, input image $\mathbf{x}_{0}$, \dots
    \Ensure purified image $\hat{\mathbf{x}}_{0}$
    \State // diffusion process
    \For {each round t=$1, \dots, T$}
      \State calculate diffused image $\mathbf{x}_{t}$ with E.q.~\ref{eq_ddpm_diffusion_alpha} and E.q.~\ref{eq_ddpm_reparameter}
    \EndFor
    \State // reverse process
    \For {each round t=$T, \dots, 1$}
      \State $\bm{\mu}, \Sigma \leftarrow \bm{\mu_{\mathbf{w}}}(\hat{\mathbf{x}}_{t},t), \Sigma_{\mathbf{w}}(\hat{\mathbf{x}}_{t},t)$ 
      \State sample $\hat{\mathbf{x}}_{t-1}$ from $\mathcal{N} (\bm{\mu} + \lambda\Sigma \nabla_{\hat{\mathbf{x}}_{t}}\left\|\mathbf{x}_{t} - \hat{\mathbf{x}}_{t} \right\|, \Sigma\mathbf{I} )$
    \EndFor
    \end{algorithmic}
\end{algorithm}

\subsection{Adversarial Image Detection}
The image purification function guarantees that all uploaded images used for recommendations are unlikely to contain adversarial perturbations. 
However, in real-life scenarios, it is also essential to detect adversarial images as it provides insights into system maintenance.
For example, the system manager can collect detected images to analyze potential attacks and even punish the detected adversarial image providers directly.

Detecting adversarial images in FedRecs is non-trivial since we only have normal images and we cannot get adversarial images for training before we can detect them.
As we know, image poisoning attacks achieve adversarial goals by adding imperceptible perturbations to the image.
These perturbations can cause remarkable changes for the image feature vectors which are encoded by the extractor $\bm{\Phi}$.
Based on this characteristic, we propose a training-free method to detect adversarial images.
Specifically, we assume that the image purified by Algorithm~\ref{alg_ddpm_guidance} will be a ``safe'' image.
In other words, the purified image will not cause remarkable changes to its encoding feature, since the purifications are destroyed.
Therefore, we employ the feature extractor to encode the image before and after purification and compare the difference between these two feature vectors.
If the similarity of these two feature vectors is smaller than a threshold $\rho$, the image will be detected as an adversarial image:
\begin{equation}
  \label{eq_detector}
  \begin{aligned}
    cos\_sim(\bm{\Phi}(\mathbf{x}_{0}) - \bm{\Phi}(\hat{\mathbf{x}}_{0})) < \rho
  \end{aligned}
\end{equation}
We use cosine similarity to measure the difference of image vectors\footnote{We also tried Euclidean distance and get equivalent experimental results.}.
$\rho$ is a preset hyper-parameter and its value will directly influence the accuracy of the detector.
In this paper, we set $\rho$ as follows:
first, we use our DDPM to purify a large number of clean images, such as those from public datasets like ImageNet. Next, we calculate the difference between purified and original images and use statistics to define $\rho$. If the difference between purified and original images is smaller than $\rho$, such an image is likely different from clean images from the image extractor perspective, indicating that it is an adversarial image.


\section{Experiments} \label{sec_exp}
In this section, we take extensive experiments to answer the following research questions (RQs):
\begin{itemize}
  \item \textbf{RQ1.} Are current state-of-the-art model poisoning attacks still effective for visually-aware federated recommender systems?
  \item \textbf{RQ2.} Are there any risks of using visual information in federated recommender systems? i.e., Are our PSMU(V) and PSMU++ effective for visually-aware federated recommendations?
  \item \textbf{RQ3.} How is the effectiveness of our diffusion model-based defense method?
\end{itemize}

\subsection{Dataset}
In this paper, we leverage two popular federated recommendation datasets for evaluation: MovieLens-1M (ML)~\cite{harper2015movielens} and Amazon Cell Phone (AZ)~\cite{mcauley2015image}.
ML contains $6,040$ users and $3,706$ items with $1,000,208$ feedback, and $3,301$ items have image descriptions.
AZ includes $103,593$ interactions with $13,174$ users, $5,970$ products, and $5,877$ visual signals.
All users have at least $5$ interactions with different products.
Following~\cite{yuan2023manipulating,zhang2022pipattack}, we binarize the user-item ratings, where all ratings are transformed to $r_{ij}=1$ and negative instances are sampled with $1:4$ ratio.
Table~\ref{tb_dataset} illustrates the basic statistics of these two datasets. 
It is worth pointing out that we choose two datasets with very different data sparsity to show the data sparsity problem's impacts on our poisoning attacks. 

\begin{table}[!ht]
  \centering
  \setlength\tabcolsep{2.pt}
  \caption{Statistics of recommendation datasets}\label{tb_dataset}
  \begin{tabular}{lccccc}
  \hline
  \textbf{Dataset}  & \multicolumn{1}{l}{\textbf{\#users}} & \multicolumn{1}{l}{\textbf{\#items}} & \multicolumn{1}{l}{\textbf{\#interactions}} & \multicolumn{1}{l}{\textbf{Avg.}} & \multicolumn{1}{l}{\textbf{Density}} \\ \hline
  ML    & 6,040                                  & 3,706                                 & 1,000,208                                     & 166                                      & 4.46\%                               \\
  AZ  & 13,174                               & 5,970                                & 103,593                                     & 8                                        & 0.13\%                               \\ \hline
  \end{tabular}
  \end{table}

\subsection{Evaluation Protocol}
For both model poisoning attacks and image poisoning attacks, the evaluation protocol is consistent with our previous work~\cite{yuan2023manipulating}.
Specifically, FedRecs are trained without attacks for a few epochs and then we launch the attacks.
The FedRecs are trained until convergence or reaching the pre-defined maximum global epochs.
We select the most unpopular items as target items.
ER@5 is used to evaluate both model poisoning attacks and image poisoning attacks.
The purification of the defense method is evaluated from two aspects: (1) whether it can reduce target items' ER@5 to normal level; (2) whether it deteriorates recommendation performance (NDCG@20).
The detection of the defense method is evaluated by accuracy.


\subsection{Parameter Settings}
All the experiments are implemented using PyTorch~\cite{paszke2019pytorch}.
For both Fed-NCF, Fed-VNCF, Fed-LightGCN, and Fed-LightVGCN, the dimension of user and item embeddings are set to $32$ following~\cite{yuan2023manipulating}.
We use the deep pretrained CNN model from~\cite{simonyan2014very} as our visual extractor $\bm{\Phi}$\footnote{We also tried other visual extractors such as ResNet and get similar results and trends, since our attack and defense do not make any special assumption on the visual extractor.}.
Then, the visual feature extracted by CNN model is transformed to $32$ dimension size vector by the visual feature transform matrix $\mathbf{E}$.
Three layers of feedforward layers are utilized to process the concatenated user, item, and visual feature vectors (optional) with sizes of $[96, 32, 16]$ for Fed-VNCF and Fed-LightVGCN and $[64, 32, 16]$ for Fed-NCF and Fed-LightGCN, respectively.
The layer of LightGCN propagation is $1$ for both Fed-LightGCN and Fed-LightVGCN.
Adam~\cite{kingma2014adam} with a learning rate of $0.001$ is adopted as the optimizer.
The same as in~\cite{yuan2023manipulating}, all poisoning attacks are launched at $8$th global epoch.
The number of selected items for synthetic users $\left|\widetilde{\mathcal{V}}_{i}^{+}\right|$ is $30$.
For model poisoning attacks, the proportion of malicious users $\xi$ equals $0.1\%$ without specific mention.
For image poisoning attacks, $\epsilon$ is $4$ as default, which is much smaller than visual attacks in centralized recommendation~\cite{liu2021adversarial,cohen2021black}.
For the diffusion model, we use an unconditional $256*256$ DDPM\footnote{\url{https://openaipublic.blob.core.windows.net/diffusion/jul-2021/256x256_diffusion.pt}} pretrained by~\cite{dhariwal2021diffusion}.
The number of diffusion steps is set to $1000$ according to \cite{dhariwal2021diffusion}.
The guidance factor $\lambda$ is $1000$.

\subsection{Effectiveness of Model Poisoning Attacks for Visually-aware FedRecs (RQ1)}\label{sec_rq1}
\begin{figure}[t]
  \centering
  \includegraphics[width=\linewidth]{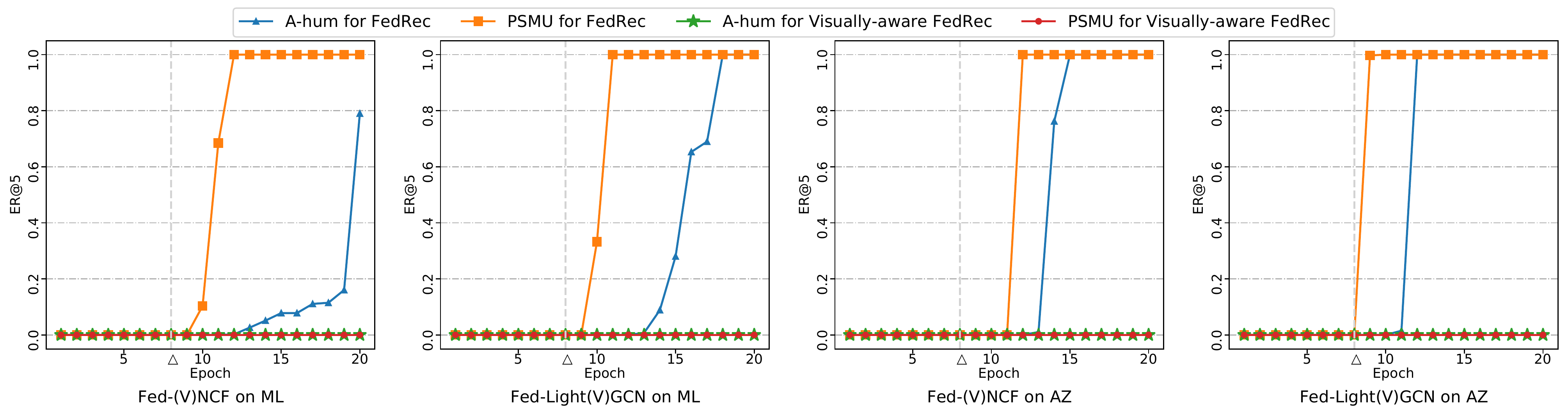}
  \caption{The performance comparison of existing state-of-the-art (SOTA) model poisoning attacks for general federated recommender systems and visually-aware federated recommender systems. All current SOTA model poisoning attacks are ineffective in visually-aware FedRecs.}\label{fig_model_poison_vfr}
  \Description{Overview of image poisoning attack.}
\end{figure}

All existing model poisoning attacks~\cite{zhang2022pipattack,rong2022fedrecattack,rong2022poisoning}, including our previous work~\cite{yuan2023manipulating}, have demonstrated their effectiveness only in FedRecs with collaborative data.
In this paper, we argue that the effectiveness of these attacks is due to the sparsity of collaborative information, which results in less robust item embeddings (especially for cold items) due to insufficient data description. Therefore, when additional item auxiliary information such as product visual description is incorporated, these poisoning attacks may become ineffective.

To support our argument, we conduct experiments with both general FedRecs and visually-aware FedRecs using A-hum and PSMU.
A-hum is the earlier state-of-the-art model poisoning attack proposed by Rong et al~\cite{rong2022poisoning}.
PSMU is the current state-of-the-art model poisoning attack proposed by our previous work~\cite{yuan2023manipulating}.
We choose these two attacks to do experiments since our previous work~\cite{yuan2023manipulating} already showed that other model poisoning attacks have very poor performance with limited malicious users ($\xi=0.1\%$).

Fig.~\ref{fig_model_poison_vfr} presents the performance comparison of these two model poisoning attacks on FedRecs with or without visual information.
For Fed-NCF and Fed-LightGCN, both PSMU and A-hum have the ability to influence the value of exposure rate on all datasets, meanwhile, our PSMU achieves better performance than A-hum (i.e., achieving higher ER@5 values or achieving ER@5=1.0 with fewer epochs.)
This result proves the effectiveness of these two model poisoning attacks for FedRecs with only collaborative information.
Besides, by comparing the same attack's performance in the same FedRec across datasets, we can observe that the sparser the dataset is, the better performance the attack has.
For example, PSMU obtains $1.0$ ER@5 scores using about $4$ and $3$ epochs on the ML dataset for Fed-NCF and Fed-LightGCN, while it only costs $3$ and $1$ epochs on AZ.
However, when FedRecs were equipped with product visual information, all these attacks' ER@5 scores dropped to $0$ in all cases (the line of ``A-hum for Visually-aware FedRec'' and ``PSMU for Visually-aware FedRec'' in Fig.~\ref{fig_model_poison_vfr}), which indicates that incorporating visual information can make Fed-NCF and Fed-LightGCN more robust to malicious poisoning attacks.

\begin{figure}[t]
  \centering
  \subfloat[Fed-VNCF on ML.]{\includegraphics[width=1.37in]{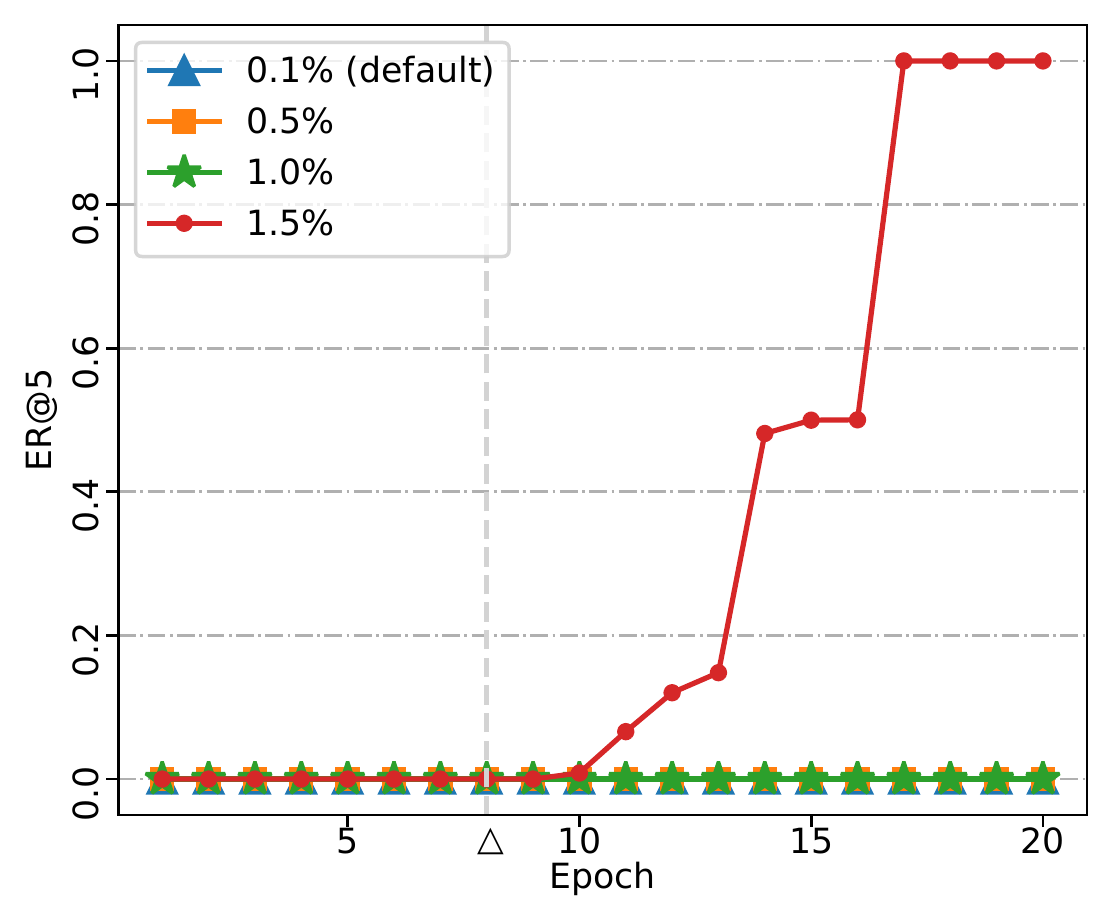}\label{fig_ml_vncf_mp_prop}}
  \subfloat[Fed-VNCF on AZ.]{\includegraphics[width=1.37in]{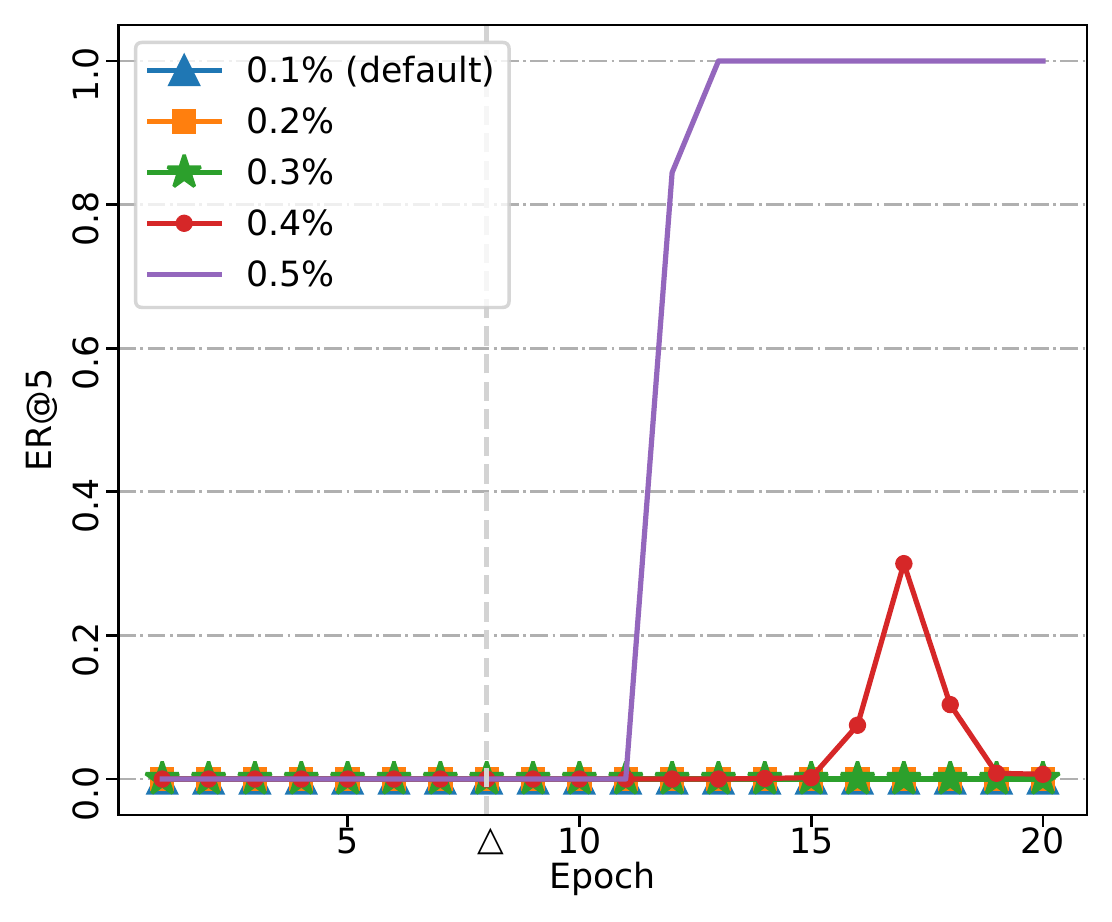}\label{fig_az_vncf_mp_prop}}
  \subfloat[Fed-LightVGCN on ML.]{\includegraphics[width=1.37in]{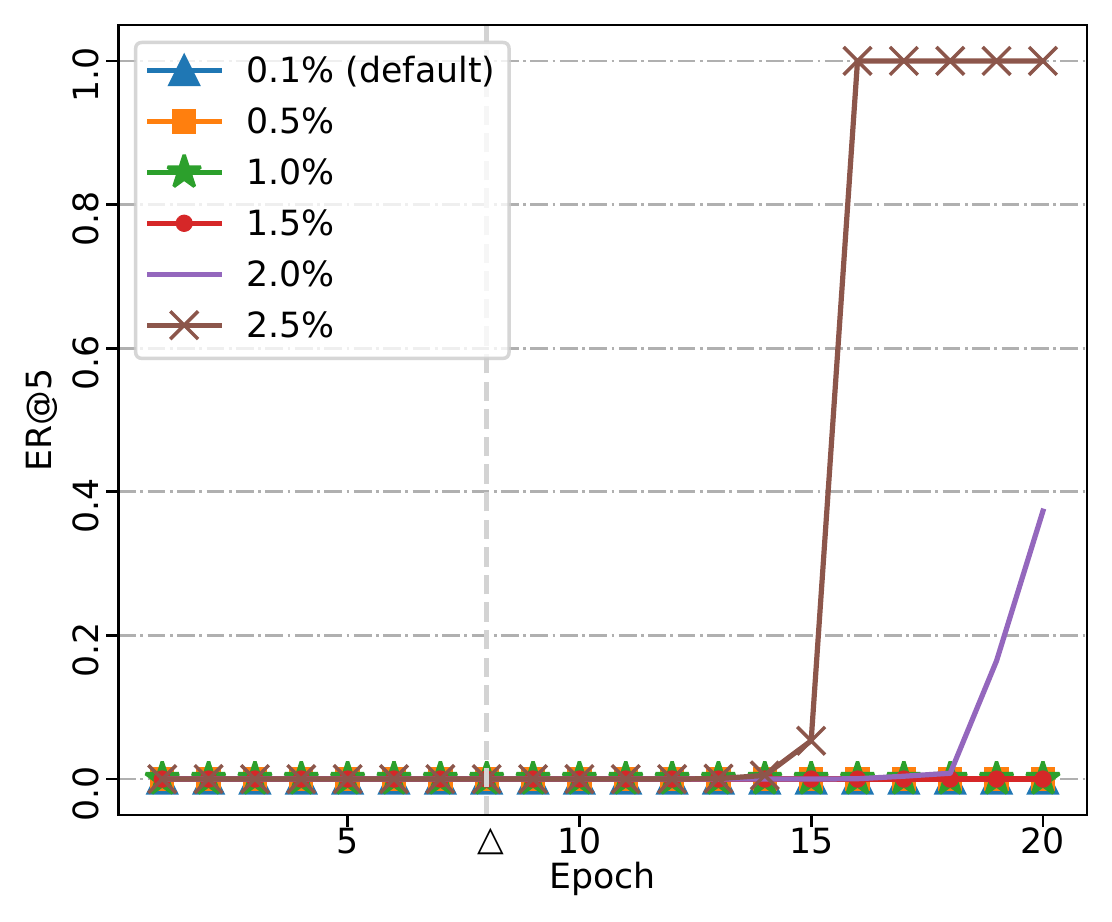}\label{fig_ml_vgcn_mp_prop}}
  \subfloat[Fed-LightVGCN on AZ.]{\includegraphics[width=1.37in]{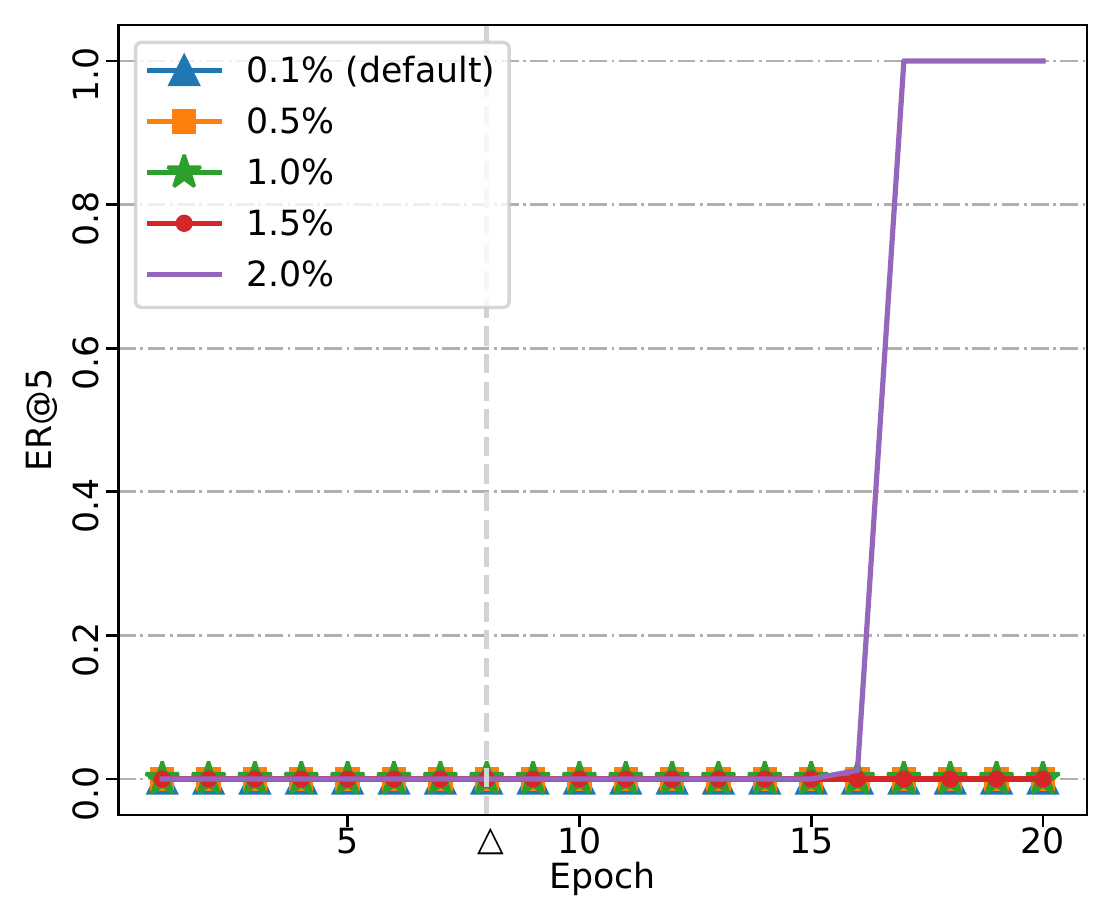}\label{fig_az_vgcg_mp_prop}}
  \caption{The effectiveness of PSMU with more malicious user proportion in visually-aware FedRecs.}\label{fig_mp_prop}
  \Description{Model deviation}
\end{figure}
In Fig.~\ref{fig_model_poison_vfr}, we have already demonstrated that PSMU, our state-of-the-art model poisoning attack, is unable to promote target items in visually-aware FedRecs when using the default setting of our previous work~\cite{yuan2023manipulating}.
However, the effectiveness of PSMU may increase with more malicious users, although the cost of launching such attacks will also increase. 
Therefore, we investigate the effectiveness of PSMU with more malicious users in Fig.~\ref{fig_mp_prop}.
The results show that PSMU requires a higher proportion of malicious users to be effective in visually-aware FedRecs. 
Specifically, PSMU cannot be effective until the proportion of malicious users increased to $1.5\%$ and $0.5\%$ for Fed-VNCF on ML and AZ, respectively, which are $15$ and $5$ times higher than the settings for FedRecs with only collaborative data.
To manipulate Fed-LightVGCN, PSMU requires at least $25$ and $20$ times more malicious users on ML and AZ than the default settings.
As a result, the costs of utilizing model poisoning attacks to compromise visually-aware FedRecs are much higher than the original FedRecs.
Furthermore, by comparing different models on the same dataset (i.e., Fig.~\ref{fig_ml_vncf_mp_prop} and Fig.~\ref{fig_ml_vgcn_mp_prop}, Fig.~\ref{fig_az_vncf_mp_prop} and Fig.~\ref{fig_az_vgcg_mp_prop}), 
we find that Fed-LightVGCN is relatively more robust than Fed-VNCF when facing model poisoning attacks.
This is because visual information is fully utilized in Fed-LightVGCN compared to Fed-NCF: Fed-LightVGCN not only uses visual information for directly predicting the preference scores (E.q.~\ref{eq_ncf}) but also utilizes it during LightGCN propagation (E.q.~\ref{eq_lightgcn}).

By combining Fig.~\ref{fig_model_poison_vfr} and Fig.~\ref{fig_mp_prop} we can conclude that incorporating visual information can improve the robustness of FedRecs for current state-of-the-art model poisoning attacks.

\subsection{Effectiveness of PSMU(V) and PSMU++ (RQ2)}
Although Section~\ref{sec_rq1} manifests that using visual information can defend against model poisoning attacks, in this subsection, we disclose that visual information will create new backdoors for adversaries to promote items by presenting the effectiveness of PSMU(V) (RQ2).
Besides, we further reveal that the backdoor of visual information gives adversaries an opportunity to simultaneously launch image and model poisoning attacks (PSMU++) to manipulate item ranks.

As mentioned in Section~\ref{sec_intro}, we are the first to present image poisoning attacks in visually-aware FedRecs.
Most previous visual attacks~\cite{liu2021adversarial,cohen2021black} in the centralized recommendation are not applicable in FedRecs settings since they depend on the feedback of benign users.
For comparison purposes, we construct the following baselines: No Attack and Popularity Attack. No Attack displays the original exposure rate of target items.
Popularity Attack is similar to the EXPA attack proposed by~\cite{liu2021adversarial}, but it differs in that it gradually changes item images during the training process. In Popularity Attack, we assume that adversaries have knowledge of the popularity information of items. At each global epoch, the attacker tries to add noise to make the target item's visual vector close to the feature vector of popular items.

\begin{figure}[t]
  \centering
  \includegraphics[width=\linewidth]{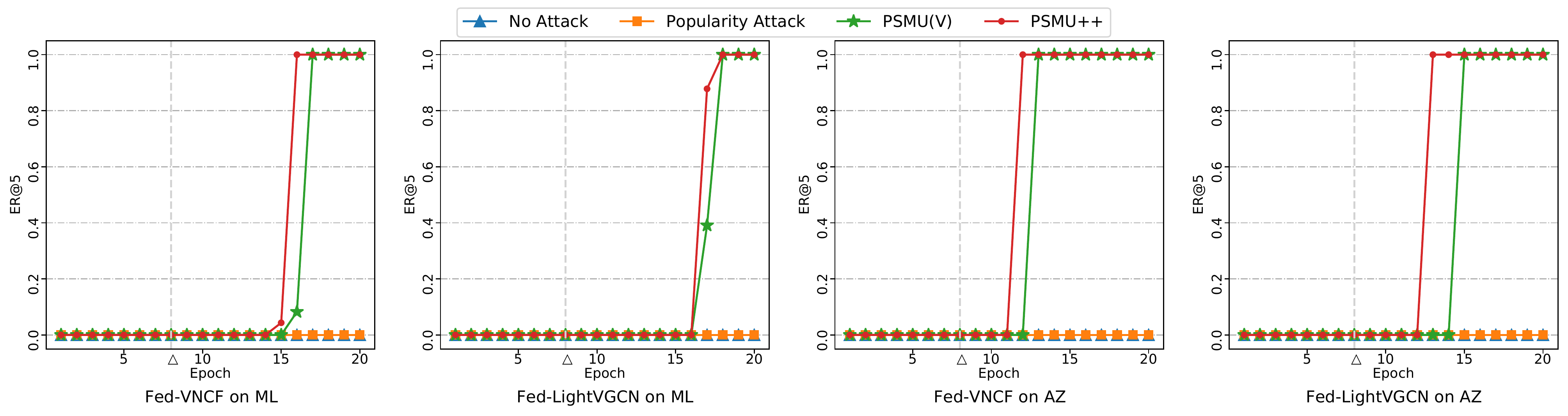}
  \caption{The trend of exposure rate of target items for different image poisoning attacks with $\epsilon=4$.}\label{fig_psmuv}
  \Description{Overview of image poisoning attack.}
\end{figure}

\begin{figure}[!htbp]
  \centering
  \includegraphics[width=\linewidth]{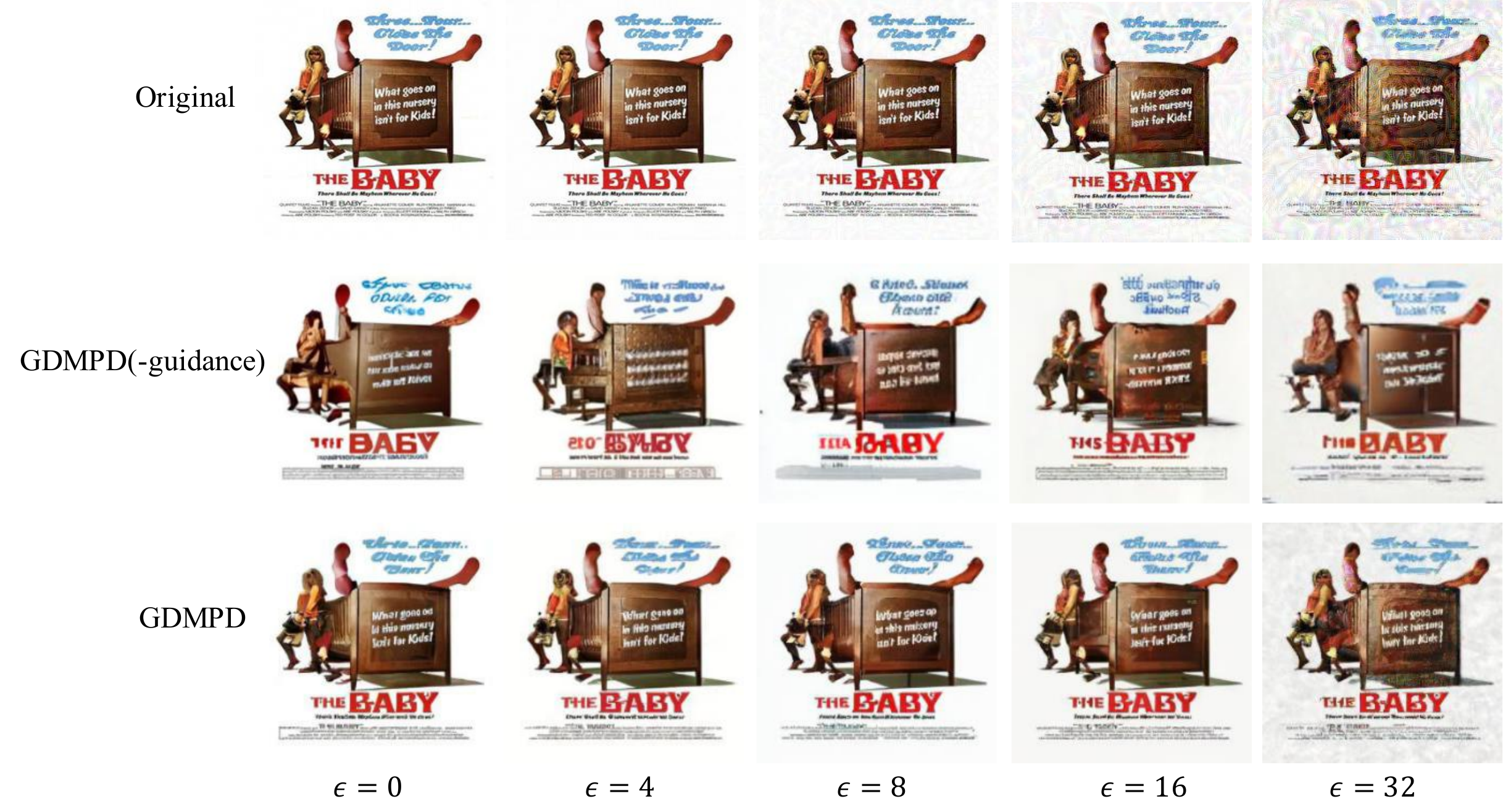} 
  \caption{Example of adversarial images with different perturbation constraints ($\epsilon$) and corresponding purified images generated by different purification methods. All these adversarial images are generated by PSMU(V) and can achieve ER@5=1.0. The original image is from the ML dataset.}\label{fig_noise_example}
  \Description{Overview of image poisoning attack.}
\end{figure}

Fig.~\ref{fig_psmuv} presents the results of different image poisoning attacks. 
Both PSMU(V) and PSMU++ use $0.1\%$ synthetic users which is the same as the setting in model poisoning attacks.
In Fig.~\ref{fig_psmuv}, we can see that No Attack and Popularity Attack cannot create any changes in the exposure rate of the target item. In contrast,
PSMU(V) promotes target items to all users with $9$, $10$, $5$, and $7$ global epochs in the cases of ``Fed-VNCF on ML'', ``Fed-LightVGCN on ML'', ``Fed-VNCF on AZ'', and ``Fed-LightVGCN on AZ'', respectively.
Moreover, when incorporating PSMU (i.e., PSMU++), the item promotion process is accelerated as shown by the red line in Fig.~\ref{fig_psmuv}.
Besides, comparing the results from different datasets, we can get a consistent conclusion with Fig.~\ref{fig_model_poison_vfr}: Promoting items is easier on AZ than on ML since AZ is sparser.
Additionally, Fed-LightVGCN is relatively more reliable than Fed-NCF under poisoning attacks since the visual information is fused by not only concatenation but also with LightGCN propagation.
It is worth noting that we set the normalization of perturbations to be less than $4$, making the polluted image be human-imperceptible.
The first line of Fig.~\ref{fig_noise_example} provides an example of adversarial images with different perturbation scales. 
$\epsilon=0$ represents the original image. 
The adversarial image with $\epsilon =4$ perturbations is indistinguishable from the original image to humans, ensuring the stealthiness of our image poisoning attacks.

\subsection{The Effectiveness of GDMPD}
The effectiveness of PSMU(V) and PSMU++ reveals the backdoor created by incorporating visual information from external sources.
In this paper, we propose a safe way to utilize images from untrustworthy sources through GDMPD, which can purify images and detect adversarial images. In this section, we conduct experiments to demonstrate the effectiveness of our GDMPD. First, we show the purification effectiveness, followed by the accuracy of GDMPD detection.
\begin{figure}[!htbp]
  \centering
  \includegraphics[width=\linewidth]{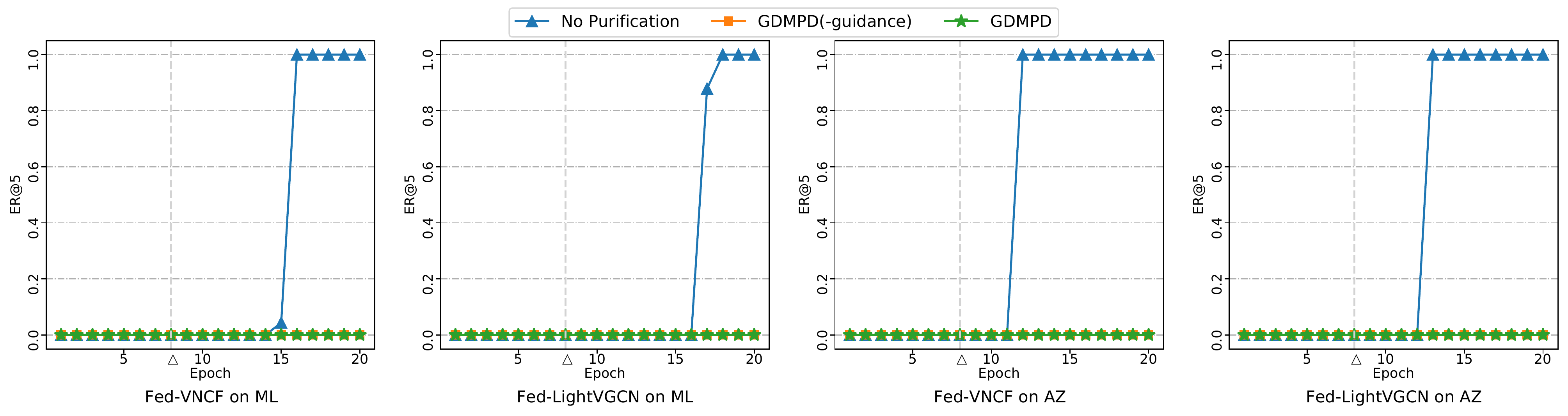}
  \caption{The trend of exposure rate of PSMU++ for FedRecs with purification mechanism.}\label{fig_purified}
  \Description{Overview of image poisoning attack.}
\end{figure}

\begin{table}[!htbp]
  \centering
  \caption{The comparison of recommendation performance (NDCG@20) for different purified methods in visually-aware FedRecs.}\label{tb_rec_perform_purify}
  \begin{tabular}{l|l|c|c}
  \hline
  \textbf{FedRec Model}                  & \textbf{Purification Method} & \textbf{ML} & \textbf{AZ} \\ \hline
  \multirow{3}{*}{\textbf{Fed-VNCF}}      & \textbf{original}            &    0.03985         &   0.02786          \\
                                         & \textbf{GDMPD(-guidance)}    &    0.04833         &   0.02806          \\
                                         & \textbf{GDMPD}               &    \textbf{0.05032}        &    \textbf{0.02849}        \\ \hline
  \multirow{3}{*}{\textbf{Fed-LightVGCN}} & \textbf{original}            &    0.03831         &   0.02065          \\
                                         & \textbf{GDMPD(-guidance)}    &    0.04494         &   0.02135          \\
                                         & \textbf{GDMPD}               &    \textbf{0.04623}         &   \textbf{0.02246}          \\ \hline
  \end{tabular}
  \end{table}

  \begin{table}[!htbp]
    \caption{The standard deviation of Blur and Brisque scores of images with different purification methods. Lower values indicate that the image quality difference is less.}\label{tb_img_quality}
    \begin{tabular}{l|cc|cc}
    \hline
                                 & \multicolumn{2}{c|}{\textbf{ML}}   & \multicolumn{2}{c}{\textbf{AZ}}    \\ \hline
    \textbf{Purification Method} & \textbf{Blur}   & \textbf{Brisque} & \textbf{Blur}   & \textbf{Brisque} \\ \hline
    \textbf{original}            & 2444.68         & 11.82            & 1682.55         & 18.138           \\
    \textbf{GDMPD(-guidance)}    & 821.41         & 8.83             & 552.84          & 16.80            \\
    \textbf{GDMPD}               & \textbf{801.02} & \textbf{8.62}    & \textbf{523.75}  & \textbf{16.59}   \\ \hline
    \end{tabular}
    \end{table}

  \begin{table}[!htbp]
    \caption{The effectiveness of GDMPD for defending against PSMU++ with different perturbation scales. The value of each cell $(x,y)$ represents: $x$ is the highest ER@5 scores that PSMU++ achieves in FedRecs without defense method, and $y$ is the highest ER@5 scores that PSMU++ achieves when equipped with GDMPD.}\label{tb_epsilon}
    \begin{tabular}{l|llll|llll}
    \hline
    \textbf{Dataset} & \multicolumn{4}{c|}{\textbf{Fed-VNCF}} & \multicolumn{4}{c}{\textbf{Fed-VLightGCN}} \\ \hline
    $(x,y)$        & $\epsilon=4$     & $\epsilon=8$     & $\epsilon=16$    & $\epsilon=32$    & $\epsilon=4$        & $\epsilon=8$        & $\epsilon=16$        & $\epsilon=32$       \\ \hline
    \textbf{ML}      & (1.0, 0.0)      & (1.0, 0.0)      &   (1.0, 0.0)    &  (1.0, 0.0)     &  (1.0, 0.0)        & (1.0, 0.0)         & (1.0, 0.0)          &  (1.0, 0.0)        \\ \hline
    \textbf{AZ}      & (1.0, 0.0)      &   (1.0, 0.0)    & (1.0, 0.0)      & (1.0, 0.0)      & (1.0, 0.0)         & (1.0, 0.0)         & (1.0, 0.0)          & (1.0, 0.0)         \\ \hline
    \end{tabular}
    \end{table}

To safely use external images, we incorporate purification mechanisms in visually-aware FedRecs. 
Fig.~\ref{fig_purified} shows PSMU++'s attack results for different purification methods.
GDMPD(-guidance) is GDMPD without diffused image guidance.
In Fig.~\ref{fig_purified}, all purification methods can reduce the attack's ER@5 to $0$, which demonstrates that by adding Gaussian noises during the diffusion process, the perturbations have been diluted.
An effective defense method should not only prevent the attacker's achieving its malicious goals, but also consume less recommendation performance.
Table~\ref{tb_rec_perform_purify} presents the recommendation performance of visually-aware FedRecs with different purification methods.
``original'' is the visually-aware FedRecs that leverage original images. 
According to the results in Table~\ref{tb_rec_perform_purify}, FedRecs with purified images even have better performance than using original images.
This is because the diffusion model can shrink the variance of original images, where images are provided by different providers and the quality of them is different.
Table~\ref{tb_img_quality} provides a proof-of-concept.
We calculate the standard deviation of Brisque~\cite{mittal2012no} and Blur scores (Laplacian operator value) to evaluate the quality deviation of images generated by different methods.
In Table~\ref{tb_img_quality}, the standard deviations of original images in both ML and AZ are much higher than purified images, indicating the original images' large quality difference.

When adding more perturbations, the attacker's goal will be easier to achieve.
Therefore, we evaluate our defense method's effectiveness with increased $\epsilon$ from $4$ to $32$ in Table~\ref{tb_epsilon}.
The results show that before utilizing our purification mechanism, the attack's ER@5 can reach to $1.0$ with all different $\epsilon$ scales.
However, when incorporating our defense methods, the scores of ER@5 reduce to $0.0$ in all cases, which implies that our defense methods can at least tolerate PSMU++ with $\epsilon<32$.
Fig.~\ref{fig_noise_example} presents a case study of adversarial images with different perturbation scales purified by different methods.
The comparison of purified images and adversarial images with large-scale perturbations (e.g., $\epsilon=16$ or $32$) shows that the diffusion model can remove abnormal noise.
Furthermore, by comparing the images generated by our GDMPD and GDMPD(-guidance), we can see that after adding guidance, the generated images are more consistent with the original ones, demonstrating the effectiveness of our guidance.

\begin{figure}[!htbp]
  \centering
  \includegraphics[width=0.45\linewidth]{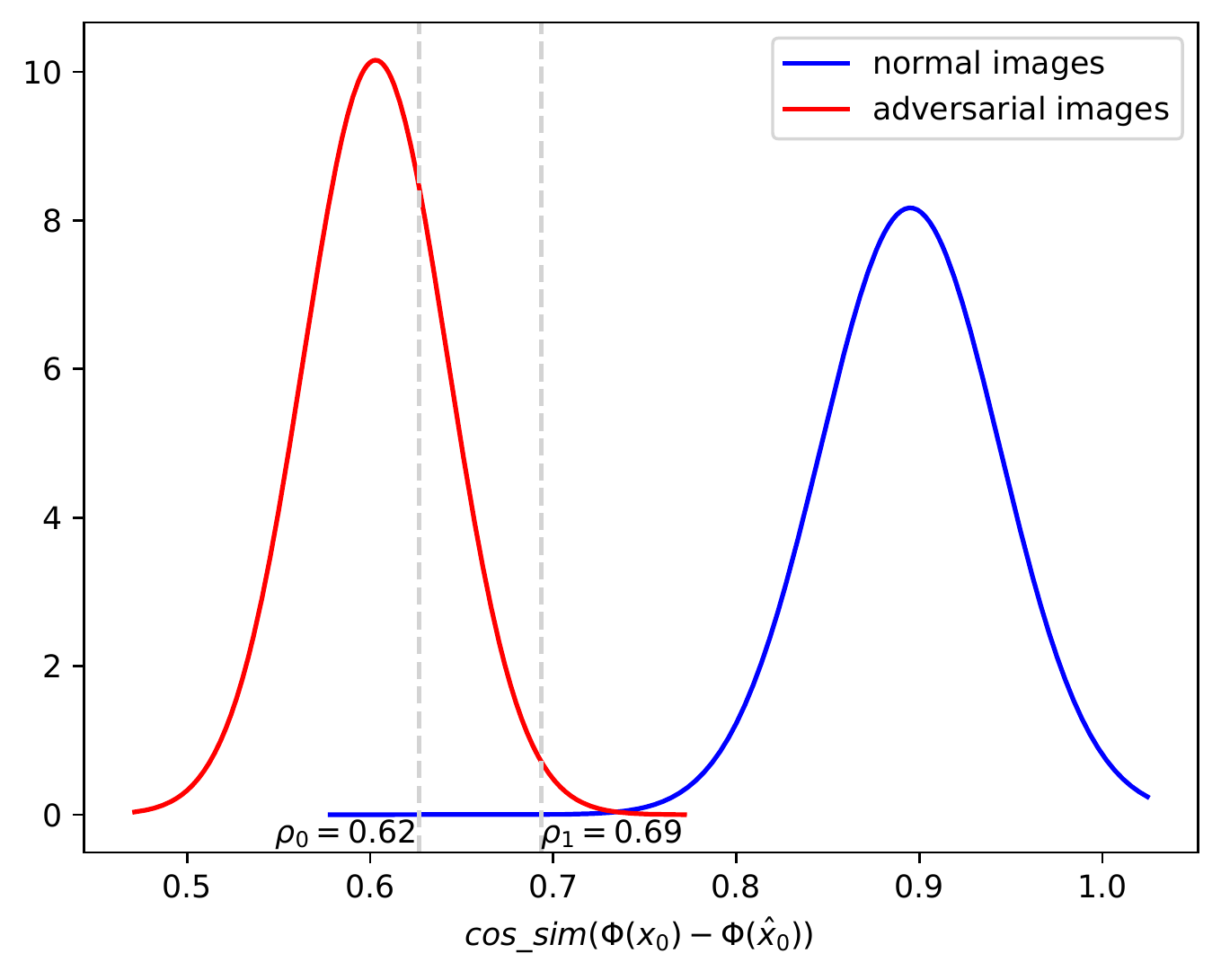}
  \caption{Visualization of the density distribution of normal and adversarial images similarity scores fitting with normal distribution. $\rho_{0}$ is the minimal similarity score from normal images. $\rho_{1}$ is the best setting of $\rho$ that can filter out all adversarial images.}\label{fig_detection}
  \Description{Overview of image poisoning attack.}
\end{figure}
To improve the maintenance of recommender systems, we have implemented a detection function in GDMPD that is based on its purification ability. 
Since our detector is training-free, the detection results for ML and AZ are almost identical, so we present the overall results for the union set of ML and AZ adversarial images. 
Specifically, we use PSMU++ to generate $100$ images that can promote cold items to $1.0$ on ML and AZ respectively. 
Then, we tested whether GDMPD can detect these adversarial images when mixed with all other normal item images. 
As described in E.q.~\ref{eq_detector}, the accuracy of adversarial image detection mainly depends on the setting of $\rho$.
In our experiments, we set $\rho$ as follows.
First, we randomly sample a subset ($10,000$ in our experiments) of images from a publicly available image dataset, ImageNet.
These images are normal images and we use our GDMPD to purify them.
After purification, we calculate the cosine similarity between purified and original images.
Finally, we naively use the minimal value of the similarity scores as $\rho$.
If the original image from FedRecs has smaller scores than $\rho$ with its corresponding purified image, GDMPD will mark it as an adversarial image.
Based on this setting, we get $0.72$ accuracy for detecting adversarial images and no normal images are falsely predicted as adversarial images.
To further analyze our detection, we visualize the distribution of normal images' similarity scores and adversarial images' similarity scores in Fig.~\ref{fig_detection}.
$\rho_{1}=0.69$ is the ``best'' setting of $\rho$ that can achieve $1.0$ accuracy to detect adversarial images from FedRecs but we cannot directly get $\rho_{1}$ in practice since we have limited prior knowledge of adversarial images.
In this paper, we simply set $\rho$ to $\rho_{0}$ according to the minimal value of normal images' similarity scores.
How to estimate a better $\rho$ can be explored in future work.

\section{Conclusion}\label{sec_conclusion}
Recently, numerous studies have exposed the threat of model poisoning attacks on federated recommender systems (FedRecs) that rely on collaborative data. We argue that these attacks are effective due to the sparsity of user-item interactions in the data. In this paper, we propose the incorporation of visual information to alleviate the data sparsity problem and demonstrate that existing model poisoning attacks cannot easily promote target items in visually-aware FedRecs. Subsequently, we propose PSMU(V) image poisoning attacks that exploit the newly created backdoor in visually-aware FedRecs. These attacks can work in tandem with model poisoning attacks, posing a greater threat and highlighting the need for a secure visual information usage mechanism. To address this gap, we propose a novel image poisoning defender based on DDPM that can not only purify adversarial images but also detect them. Extensive experiments conducted on two real-world datasets using two visually-aware FedRecs demonstrate the effectiveness of our proposed attacks and defenses.
\bibliographystyle{ACM-Reference-Format}
\bibliography{sample-base}










\end{document}